\documentclass[journal]{IEEEtran}
\usepackage{graphicx}
\graphicspath{{Pics/}}
\usepackage{epstopdf}
\usepackage{amsmath,amssymb} 
\usepackage[monochrome]{color}
\usepackage{subfigure}
\usepackage{balance}
\usepackage{enumerate}

\usepackage{enumitem}
\usepackage{booktabs}
\newcommand{\argmin}{\operatornamewithlimits{arg\,min}}
\usepackage{multirow}
\allowdisplaybreaks
 \usepackage{url}
\usepackage[numbers,sort&compress]{natbib}

\DeclareMathAlphabet{\mathbbmsl}{U}{bbm}{m}{sl}
\newcommand{\bbY}{\mathbbmsl{Y}}
\newcommand{\rmd}{{\rm d}}
\newcommand{\T}{{\top}}
\newcommand{\E}{{\rm E}}
\newcommand{\cov}{{\rm Cov}}

\newcommand{\calN}{\mathcal{N}}
\newcommand{\bbR}{\mathbbmsl{R}}

\ifCLASSINFOpdf
\else
\fi
\begin{document}
\title{Nonlinear Bayesian Estimation: From Kalman Filtering to a Broader Horizon}

\author{Huazhen~Fang,~Ning Tian,
        Yebin~Wang,~MengChu~Zhou,~and~Mulugeta~A.~Haile% <-this % stops a space
\thanks{H. Fang and N. Tian are with the Department
of Mechanical Engineering, University of Kansas, Lawrence, KS 66045, USA (e-mail: \{fang, ning.tian\}@ku.edu).}% <-this % stops a space
\thanks{Y. Wang is with the Mitsubishi Electric Research Laboratories, Cambridge, MA 02139, USA (e-mail: yebinwang@ieee.org).}
\thanks{M. Zhou is with the Department of Electrical and Computer Engineering, New Jersey Institute of Technology, Newark, NJ 07102, USA (e-mail: zhou@njit.edu).}
\thanks{M.A. Haile is with the Vehicle Technology Directorate, US Army Research Laboratory, Aberdeen, MD 21005, USA (e-mail: mulugeta.a.haile.civ@mail.mil).}
%\thanks{Manuscript received April 19, 2005; revised September 17, 2014.}
}

\vspace{-15mm}

\maketitle

\begin{abstract}
This article presents an up-to-date tutorial review of nonlinear Bayesian estimation. State estimation for nonlinear systems has been a challenge encountered in a wide range of engineering fields, attracting decades of research effort. To date, one of the most promising and popular approaches is to view and address the problem from a Bayesian probabilistic perspective, which enables estimation of the unknown state variables by tracking their probabilistic distribution or statistics (e.g., mean and covariance) conditioned on the system's measurement data. This article offers a systematic introduction of the Bayesian state estimation framework and reviews various Kalman filtering (KF) techniques, progressively from the standard KF for linear systems to extended KF, unscented KF and ensemble KF for nonlinear systems. It also overviews other prominent or emerging Bayesian estimation methods including the Gaussian filtering, Gaussian-sum filtering, particle filtering and moving horizon estimation and extends the discussion of state estimation forward to more complicated problems such as simultaneous state and parameter/input estimation.
\end{abstract}


% \fi
\IEEEpeerreviewmaketitle

\section{Introduction}
\IEEEPARstart{A}S a core subject of control systems theory, state estimation for nonlinear dynamic systems has been undergoing active research and development for a few decades. Considerable attention is gained from a wide community of researchers, thanks to its significant applications in signal processing, navigation and guidance, and econometrics, just to name a few.
When stochastic systems, i.e., systems subjected to the effects of noise, are considered, the Bayesian estimation approaches have evolved as a leading estimation tool enjoying wide popularity. Bayesian analysis traces back to the 1763 essay~\cite{Bayes:PT:1763}, published two years after the death of its author, Rev. Thomas Bayes. This seminal work was meant to tackle the following question: ``Given the number of times in which an unknown event has happened and failed: {\em Required} the chance that the probability of its happening in a single trial lies somewhere between any two degrees of probability that can be named''. Rev. Bayes developed a solution to examine the case of only continuous probability, single parameter and a uniform prior, which is an early form of the Bayes' rule known to us nowadays. Despite its preciousness, this work remained obscure for many scientists and even mathematicians of that era. The change came when the French mathematician Pierre-Simon de Laplace rediscovered the
result and presented the theorem in the complete and modern form. A historical account and comparison of Bayes' and Laplace's work can be
found in~\cite{Dale:AHES:1982}. From today's perspective, the Bayes' theorem is a probability-based answer to a philosophical question: How
should one update an existing belief when given new evidence~\cite{Malakoff:Science:1999}? Quantifying the degree of belief by probability, the theorem modifies the original belief by producing the probability conditioned on new evidence from the initial probability. This idea was applied in the past century from one field to another whenever the belief update question arose, driving numerous intriguing explorations. Among them, a topic of relentless interest is Bayesian state estimation, which is concerned with determining the unknown state variables of a dynamic system using the Bayesian theory.

The capacity of the Bayesian analysis to provide a powerful framework for state estimation has been well recognized now. A representative method within
the framework is the well-known Kalman filter (KF), which ``revolutionized the field of estimation ... (and) opened up many new theoretical and practical possibilities''~\cite{Jackson:AMS:2010}. KF was initially developed by using the least squares in the 1960
paper~\cite{Kalman:JBE:1960} but reinterpreted from a Bayesian perspective in~\cite{Ho:TAC:1964}, only four years after its invention. Further envisioned in~\cite{Ho:TAC:1964} was that ``the Bayesian approach offers a unified and intuitive viewpoint particularly adaptable to handling modern-day control problems''. This investigation and vision ushered a new statistical treatment of nonlinear estimation problems, laying a foundation for prosperity of research on this subject. 

In this article, we offer a systematic and bottom-to-up introduction to major Bayesian state estimators, with a particular emphasis on the KF family. We begin with outlining the essence of Bayesian thinking for state estimation problems, showing that its core is the model-based prediction and measurement-based update of the probabilistic belief of unknown state variables. A conceptual KF formulation can be made readily in the Bayesian setting, which tracks the mean and covariance of the states modeled as random vectors throughout the evolution of the system. Turning a conceptual KF into executable algorithms requires certain approximations to nonlinear systems; and depending on the approximation adopted, different KF methods are derived. We demonstrate three primary members of the KF family in this context: extended KF (EKF), unscented KF (UKF), and ensemble KF (EnKF), all of which have achieved proven success both theoretically and practically. A review of other important Bayesian estimators and estimation problems is also presented briefly in order to introduce the reader to the state of the art of this vibrant research area.

%%%%%%%%%%%%%%%%%%%%%%%%%%%%%%%%%%%%%%%%%%%%%%%%%%%%%%%%%%%%%%%%%%%%%%%%%%%%%%%%
%\section{Nonlinear Stochastic Estimation Overview} \label{sec:survey}
%as suggested by Associated editor
%%%%%%%%%%%%%%%%%%%%%%%%%%%%%%%%%%%%%%%%%%%%%%%%%%%%%%%%%%%%%%%%%%%%%%%%%%%%%%%%
%Nonlinear state estimation continues to receive significant attention. This interest has led to a range of methods being developed over the last four decades. Nonlinear estimation for system \eqref{eq:statespace} can be treated in deterministic or stochastic frameworks. One of the original stochastic nonlinear designs is the Extended Kalman Filter (EKF) which is based on a (time-varying) linearization of a nonlinear system about an estimated state trajectory \cite{SmiSchMcG62}. The EKF for a nonlinear system is in fact a standard Kalman Filter for the system's (LTV) linearization and therefore its performance can be fundamentally limited by its first order approximation. Among the improvement of the EKF is the Unscented Kalman filter (UKF).

%%%%%%%%%%%%%%%%%%%%%%%%%%%%%%%%%%%%%%%%%%%%%%%%%%%%%%%%%%%%%%%%%%%%%%%%%%%%%%%%
%probabilistic perspective
%%%%%%%%%%%%%%%%%%%%%%%%%%%%%%%%%%%%%%%%%%%%%%%%%%%%%%%%%%%%%%%%%%%%%%%%%%%%%%%%

\section{A Bayesian View of State Estimation}
We consider the following nonlinear discrete-time system:
\begin{equation}\label{sys}
\left\{
\begin{aligned}
x_{k+1} &= f(x_k)+w_k,\\
y_k &= h(x_k)+v_k,
\end{aligned}
\right.
\end{equation}
where $x_k \in \mathbb{R}^{n_x}$ is the unknown system state, and
$y_k\in  \mathbb{R}^{n_y}$ the output, with both $n_x$ and $n_y$ being positive integers. The process noise $w_k$
and the measurement noise $v_k$ are mutually independent, zero-mean
white Gaussian sequences with covariances $Q_k$ and $R_k$,
respectively. The nonlinear mappings $f: \mathbb{R}^{n_x}
\rightarrow \mathbb{R}^{n_x}$ and $h: \mathbb{R}^{n_x}
\rightarrow \mathbb{R}^{n_y}$ represent the process dynamics and the
measurement model, respectively. The system in~\eqref{sys} is assumed
input-free for simplicity of presentation, but the following results can be easily
extended to an input-driven system.

The state vector $x_k$ comprises a set of variables that fully describe the status or behavior of the system. It evolves through time as a result of the system dynamics. The process of states over time hence represents the system's behavior. Because it is unrealistic to measure the complete state in most practical applications, state estimation is needed to infer $x_k$ from the output $y_k$. More specifically, the significance of estimation comes from the crucial role it plays in the study of dynamic systems. First, one can monitor how a system behaves with state information and take corresponding actions when any adjustment is necessary. This is particularly important to ensure the detection and handling of internal faults and anomalies at the earliest phase. Second, high-performing state estimation is the basis for the design and implementation of many control strategies. The past decades have witnessed a rapid growth of control theories, and most of them, including optimal control, model predictive control, sliding mode control and adaptive control, premise the design on the availability of state information. 

While state estimation can be tackled in a variety of ways, the stochastic estimation has drawn remarkable attention and been profoundly developed in terms of both theory and applications. Today, it is still receiving continued interest and intense research effort.
From a stochastic perspective, the system in~\eqref{sys} can be viewed as a {\em generator of random vectors $x_k$ and $y_k$}. The reasoning is as follows. Owing to the initial uncertainty or lack of knowledge of the initial condition, $x_0$ can be considered as a random vector subject to variation due to chance. Then, $f(x_0)$ represents a nonlinear transformation of $x_0$, and its combination with $w_0$ modeled as another random vector generates a new random vector $x_1$. Following this line, $x_k$ for any $k$ is a random vector, and the same idea applies to $y_k$. In practice, one can obtain the sensor measurement of the output at each time $k$, which can be considered as a sample drawn from the distribution of the random vector $y_k$. For simplicity of notation, we also denote the output measurement as $y_k$ and the measurement set at time $k$ as $\bbY_k := \{y_1, y_2, \cdots, y_k\}$.
The state estimation then is to build an estimate of $x_k$ using $\bbY_k$ at each time $k$. To this end, one's interest then lies in how to capture $p(x_k|\bbY_k)$, i.e., the probability density function (pdf) of $x_k$ conditioned on $\bbY_k$. This is because $p(x_k | \bbY_k)$ captures the information of $x_k$ conveyed in $\bbY_k$ and can be leveraged to estimate $x_k$.

\begin{figure}[t]
  \center
  \includegraphics[width=\linewidth]{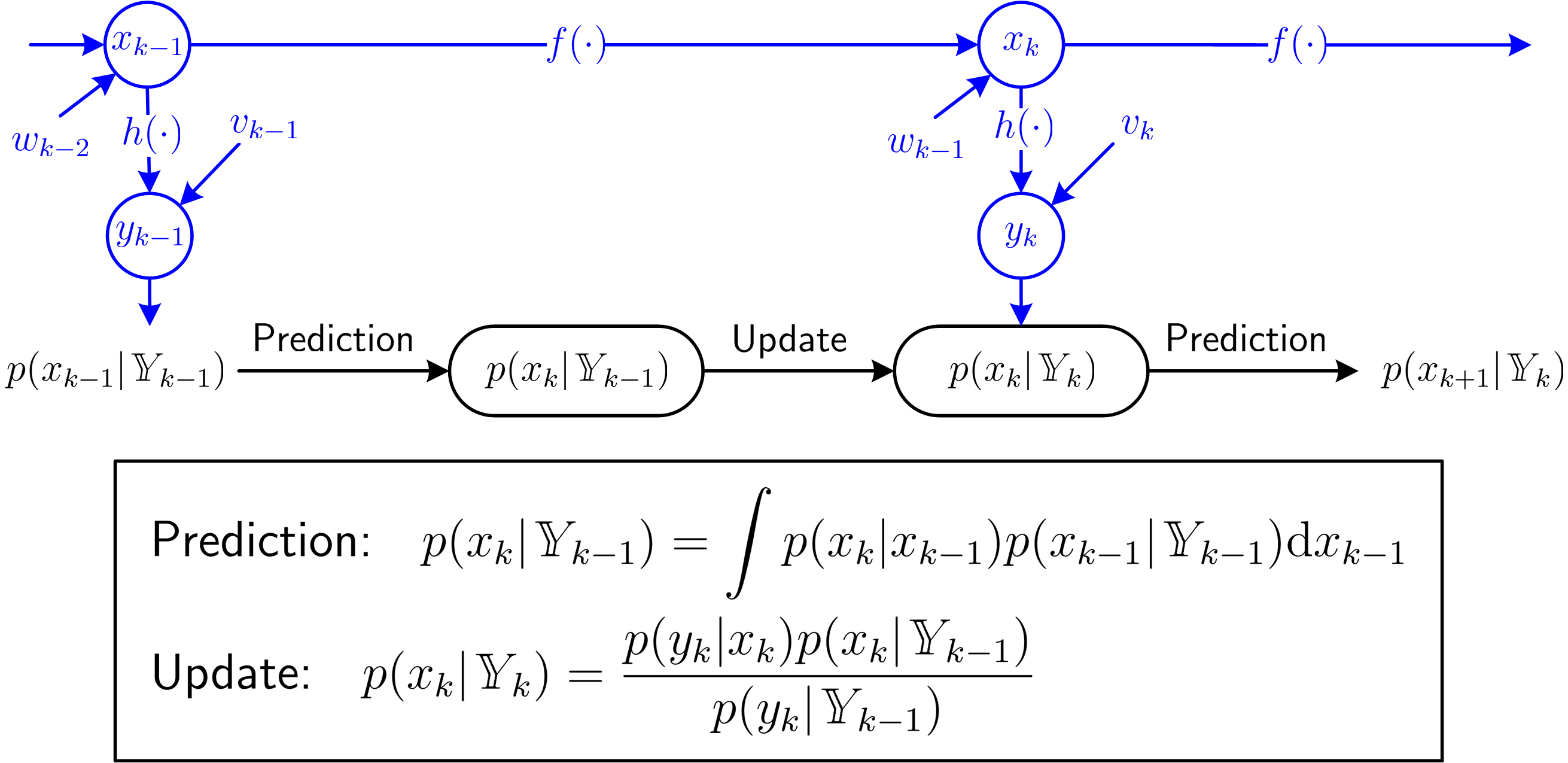}\\
  \caption{The Bayesian filtering principle. The running of a dynamic system propagates the state $x_k$ through time and produces output measurement $y_k$ at each time $k$. For the purpose of estimation, the Bayesian filtering principle tracks the pdf of $x_k$ given the measurement set $\bbY_k = \{y_1,y_2,\cdots,y_k \}$. It consists of two steps sequentially implemented: the {\em prediction} step from $p(x_{k-1}|\bbY_{k-1})$ to $p(x_{k}|\bbY_{k-1})$, and the {\em update} step from $p(x_k|\bbY_{k-1})$ to $p(x_k|\bbY_k)$ upon the arrival of $y_k$.}\label{Bayesian-estimation}
\end{figure}

A ``prediction-update'' procedure\footnote{The two steps are equivalently referred to as
`time-update' and `measurement-update', or `forecast' and `analysis', in different literature.} can be recursively executed to obtain $p(x_k|\bbY_k)$. Standing at time $k-1$, we can {\em predict} what $p(x_k|\bbY_{k-1})$ is like using $p(x_{k-1}|\bbY_{k-1})$.
When the new measurement $y_k$ conveying information about $x_k$
arrives, we can {\em update} $p(x_k|\bbY_{k-1})$ to $p(x_k|\bbY_k)$.
Characterizing a probabilistic belief about $x_k$ before and after the arrival of
$y_k$, $p(x_k | \bbY_{k-1})$ and $p(x_k | \bbY_k)$
are referred to as the {\em a priori} and {\em a posteriori} pdf's, respectively. Specifically, the prediction at time $k-1$, demonstrating the pass from $p(x_{k-1} |
\bbY_{k-1})$ to $p(x_k | \bbY_{k-1})$, is given by
\begin{align} \label{Bayesian-prediction}
p(  x_k |   \bbY_{k-1}) = \int p(  x_k |   x_{k-1} ) p(
  x_{k-1} |   \bbY_{k-1}) \rmd   x_{k-1}.
\end{align}
Let us explain how to achieve~\eqref{Bayesian-prediction}.
By the Chapman-Kolmogorov equation, it can be
seen that
\begin{align*}
  p(  x_k |   \bbY_{k-1}) = \int p\left(  x_k ,  x_{k-1} | \bbY_{k-1}\right) \rmd x_{k-1},
\end{align*}
which, according to the Bayes' rule, can be written as
\begin{align*}
  p(  x_k |   \bbY_{k-1}) =\int p(  x_k  |  x_{k-1} , \bbY_{k-1}) p(
  x_{k-1} |   \bbY_{k-1}) \rmd   x_{k-1}.
\end{align*}
It reduces to~\eqref{Bayesian-prediction}, because $p(  x_k  |
x_{k-1} , \bbY_{k-1}) =p(  x_k  |  x_{k-1} ) $ as a result of the Markovian propagation of the state.
Then on the arrival of $y_k$, $p(x_k | \bbY_{k-1})$ can be updated to
yield $p(x_k | \bbY_k)$, which is governed by
\begin{align}\label{Bayesian-update}
p(  x_k |   \bbY_k) = \frac{p(  y_k |   x_k)p(  x_k |
    \bbY_{k-1})}{p(  y_k
    |   \bbY_{k-1}) }.
\end{align}
The above equation is also owing to the use of the Bayes' rule:
\begin{align*}
p(x_k | \bbY_k) &= \frac{p(x_k, \bbY_k)}{p(\bbY_k)} =  \frac{p(x_k, y_k, \bbY_{k-1})}{p(y_k, \bbY_{k-1})} \\
&= \frac{p(y_k | x_k, \bbY_{k-1}) p(x_k, \bbY_{k-1})}{ p(y_k, \bbY_{k-1})}\\
&= \frac{p(y_k | x_k, \bbY_{k-1}) p(x_k | \bbY_{k-1})}{ p(y_k | \bbY_{k-1})}.
\end{align*}
Note that we have $p(y_k | x_k,
\bbY_{k-1}) = p(y_k | x_k)$ from the fact that $y_k$ only depends on $x_k$. Then, ~\eqref{Bayesian-update} is obtained.
Together, \eqref{Bayesian-prediction}-\eqref{Bayesian-update} represent the
fundamental principle of Bayesian state estimation for the system
in~\eqref{sys}, describing the sequential propagation of the {\em a
priori} and {\em a posteriori} pdf's. The former captures our belief
over the unknown quantities in the presence of only the prior evidence,
and the latter updates this belief using the Bayesian theory when new
evidence becomes available. The two steps, prediction
and update, are executed alternately through time, as illustrated in
Fig.~\ref{Bayesian-estimation}.

Looking at the above Bayesian filtering principle, we can summarize three elements that constitute the thinking of Bayesian estimation. First, all the unknown quantities or uncertainties
in a system, e.g., state, are viewed from the probabilistic perspective. In other words, any unknown variable is regarded as a random variable. Second, the output measurements of the system are samples drawn from a certain probability distribution dependent on the concerned variables. They provide data evidence for state estimation. Finally, the system model represents transformations that the unknown and random state variables  undergo over
time. Originating from the philosophical abstraction that anything unknown, in one's mind, is subject to variations due to chance, the randomness-based representation enjoys universal applicability even when the unknown or uncertain quantities are not necessarily random in physical sense. In addition, it can easily translate into a convenient `engineering' way for estimation of the unknown variables, as will be shown in the following discussions.

\section{From Bayesian Filtering to Kalman Filtering}\label{Bayesian-KF}

In the above discussion, we have shown the probabilistic nature of  state estimation and presented the Bayesian filtering principle~\eqref{Bayesian-prediction}-\eqref{Bayesian-update}  as a solution framework. However, this does not mean that one can simply use~\eqref{Bayesian-prediction}-\eqref{Bayesian-update} to track the conditional pdf of a random vector passing through nonlinear transformations, because the nonlinearity  often makes it difficult or impossible to derive an exact or closed-form solution. 
This challenge turns against the development of executable state estimation algorithms, since a dynamic system's state propagation and observation are based on the nonlinear functions of the random state vector $x_k$, i.e., $f(x_k)$ and $h(x_k)$. Yet for the sake of estimation, one only needs the statistics (mean and covariance) of $x_k$ conditioned on the measurements in most circumstances, rather than a full grasp of its conditional pdf. A straightforward and justifiable way is to use the mean as the estimate of $x_k$ and the covariance as the confidence (or equivalently, uncertainty) measure. Reducing the pdf tracking to the mean and covariance tracking can significantly mitigate the difficulty in the design of state estimators. To simplify the problem further, certain Gaussianity approximations can be made because of the mathematical tractability and statistical soundness of Gaussian distributions (for the reader's convenience, several properties of the Gaussian distribution to be used next are summarized in the Appendix.). Proceeding in this direction, we can reach a formulation of the Kalman filtering (KF) methodology, as shown below.

%For state prediction, the following two assumptions are made:
%\begin{asmp}
%$p(x_k | \bbY_{k})$ is a Gaussian distribution with mean $\hat
%x_{k|k}$ and $P_{k|k}^x$.
%\end{asmp}
%\begin{asmp}
%$p(x_k | x_{k-1})$ is a Gaussian distribution with mean
%$f(x_{k-1)}$ and $Q_{k-1}$.
%\end{asmp}

In order to predict $x_k$ at time $k-1$, we consider the minimum-variance unbiased estimation, which gives that the best estimate of $x_k$ given $\bbY_{k-1}$, denoted as $\hat x_{k|k-1}$, is $\E(x_k | \bbY_{k-1})$~\cite[Theorem 3.1]{Anderson:1979}. That is,
\begin{align}\label{Prediction_1}
\hat x_{k|k-1} &= \E(x_k | \bbY_{k-1}) = \int x_k p(x_k|\bbY_{k-1}) \rmd x_k.
\end{align}
Inserting~\eqref{Bayesian-prediction} into the above equation, we have
\begin{align}\label{Prediction_2}\nonumber
\hat x_{k|k-1} &= \int \left[\int x_k p(x_k|x_{k-1}) dx_k\right]\\ & \quad \quad  \quad \cdot p(x_{k-1}|\bbY_{k-1}) \rmd x_{k-1}.
\end{align}
By assuming that $w_k$ is a white Gaussian noise independent of $x_k$, we have $x_k|x_{k-1}\sim \calN(f(x_{k-1}), Q)$ and then $\int x_k p(x_k|x_{k-1}) dx_k = f(x_{k-1})$ according to~\eqref{GRV_1}. Hence,~\eqref{Prediction_2} becomes
\begin{align}\label{GF_P_1}\nonumber
\hat x_{k|k-1} &= \int f(x_{k-1}) p(x_{k-1}|\bbY_{k-1}) \rmd x_{k-1}\\ & = \E\left[ f \left(x_{k-1} | \bbY_{k-1} \right) \right].
\end{align}
For $\hat x_{k|k-1}$ in~\eqref{GF_P_1}, the associated prediction error covariance is
\begin{align}\label{Prediction_3}\nonumber
&P_{k|k-1}^x = \E\left[(x_k - \hat x_{k|k-1}) (x_k - \hat x_{k|k-1})^\T \right] \\& \quad= \int (x_k - \hat x_{k|k-1}) (x_k - \hat x_{k|k-1})^\T p(x_k|\bbY_{k-1}) \rmd x_k.
\end{align}
With the use of~\eqref{Bayesian-prediction} and~\eqref{GRV_1}, we can obtain
\begin{align}\label{GF_P_2}\nonumber
&P_{k|k-1}^x
 = \int x_k x_k^\T p(x_k|\bbY_{k-1}) \rmd x_k - \hat x_{k|k-1}\hat x_{k|k-1}^\T \\ \nonumber
& =\int \left[\int x_k x_k^\T p(x_k|x_{k-1}) \rmd x_k\right] p(x_{k-1}|\bbY_{k-1}) \rmd x_{k-1} \\ \nonumber & \quad\quad - \hat x_{k|k-1}\hat x_{k|k-1}^\T \\ \nonumber
&  = \int \left[f(x_{k-1}) f^\T(x_{k-1})+Q\right]  p(x_{k-1} | \bbY_{k-1}) \rmd x_{k-1} \\ \nonumber & \quad\quad -  \hat x_{k|k-1}  \hat x_{k|k-1}^\T\\ \nonumber
&=  \int \left[f(x_{k-1})-\hat x_{k|k-1}\right] \left[f(x_{k-1})-\hat x_{k|k-1}\right]^\T \\ \nonumber & \quad\quad  \cdot  p(x_{k-1} | \bbY_{k-1}) \rmd x_{k-1} +Q\\
&= \cov \left[ f(x_{k-1}) | \bbY_{k-1} \right] +Q.
%&= \E \left[\left. \left( f(x_{k-1}) - \hat x_{k|k-1} \right) \left( f(x_{k-1}) - \hat x_{k|k-1} \right)^\T \right| \bbY_{k-1} \right]+Q.
\end{align}

\begin{figure}[t]
  \center
  \includegraphics[width=\linewidth]{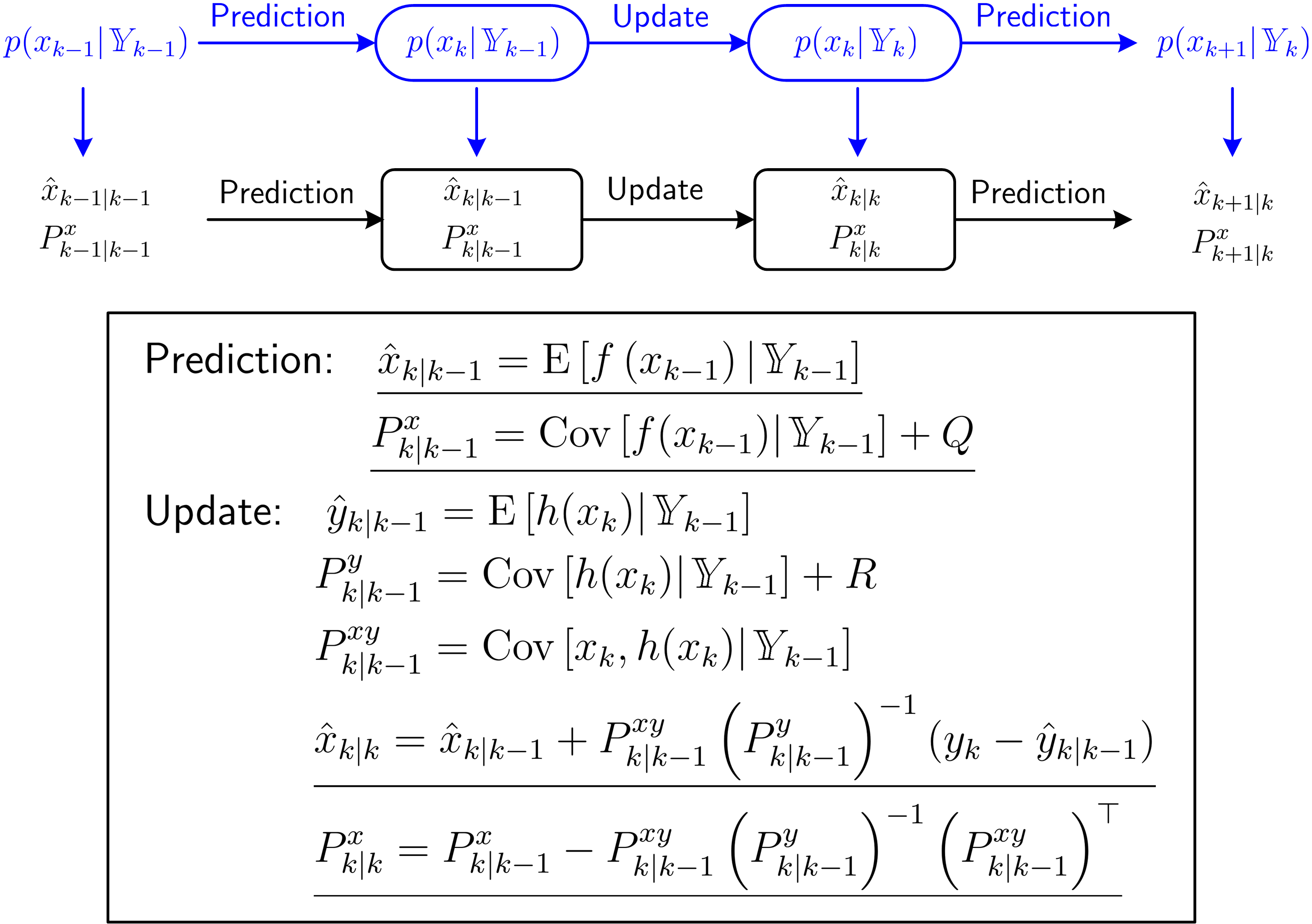}\\
  \caption{A schematic sketch of the KF technique. KF performs the prediction-update procedure recursively to track the mean and covariance of $x_k$ for estimation. The equations show that its implementation depends on determining the mean and covariance of the random state vector through nonlinear functions $f(\cdot)$ and $h(\cdot)$.}\label{Gaussian-filter}
\end{figure}

When $y_k$ becomes available, we assume that $p(x_k, y_k | \bbY_{k-1})$ can be approximated by a Gaussian distribution
\begin{align}\label{joint}
\calN\left(
\left[ \begin{matrix} \hat x_{k|k-1} \cr \hat y_{k|k-1} \end{matrix} \right],
\left[ \begin{matrix} P_{k|k-1}^x & P_{k|k-1}^{xy}\cr
 \left(P_{k|k-1}^{xy}\right)^\T & P_{k|k-1}^y \end{matrix} \right]\right),
\end{align}
where $\hat y_{k|k-1}$ is the prediction of $ y_k$ given
$\bbY_{k-1}$ and expressed as
\begin{equation}\label{y_p_temp}
\hat y_{k|k-1}  = \E ( y_k | \bbY_{k-1}) = \int y_k p(y_k|\bbY_{k-1}) \rmd y_k.
\end{equation}
The associated covariance is
\begin{align}\label{y_cov_temp}\nonumber
P_{k|k-1}^y &= \int \left(y_k-\hat y_{k|k-1} \right) \left(y_k-\hat y_{k|k-1} \right)^\T \\ & \quad\quad \cdot  p\left(y_k|\bbY_{k-1}\right) \rmd y_k.
\end{align}
It is noted that
\begin{align} \label{y_prob}\nonumber
p(y_k|\bbY_{k-1})&= \int p(x_k, y_k|\bbY_{k-1}) \rmd x_k
\\&= \int p(y_k|x_k) p  (x_k|\bbY_{k-1}) \rmd x_k.
\end{align}
Combining~\eqref{y_p_temp}-\eqref{y_cov_temp} with~\eqref{y_prob} yields
\begin{align}\label{y_p} \nonumber
\hat y_{k|k-1} &= \int \left[ \int y_k p(y_k|x_k) \rmd y_k  \right] p(x_k|\bbY_{k-1}) \rmd x_k\\ &
=\int h(x_k) p(x_k|\bbY_{k-1}) \rmd x_k= \E\left[h(x_k) | \bbY_{k-1} \right],
% \\&= \int h(x_k) \cdot \calN( x_k; \hat x_{k|k-1}, P_{k|k-1} ) \rmd x_k,
\\ \label{y_cov} \nonumber
P_{k|k-1}^y &= \int \left(h(x_k)-\hat y_{k|k-1} \right) \left(h(x_k)-\hat y_{k|k-1} \right)^\T \\ \nonumber & \quad \quad\cdot p(x_k|\bbY_{k-1})\rmd x_k+R\\
&= \cov \left[ h(x_{k}) | \bbY_{k-1} \right] +R.
\end{align}
The cross-covariance between $x_k$ and $y_k$ is
\begin{align}\label{xy_cov}  \nonumber
&P_{k|k-1}^{xy}  =  \iint \left(x_k-\hat x_{k|k-1} \right) \left(y_k-\hat y_{k|k-1}\right)^\T\\ \nonumber & \quad\quad\quad\quad\quad \cdot   p\left(x_k, y_k|\bbY_{k-1}\right) \rmd x_k \rmd y_k\\ \nonumber
&=\int \left(x_k-\hat x_{k|k-1} \right) \left[\int \left(y_k-\hat y_{k|k-1}\right)^\T   p\left(y_k|x_k\right) \rmd y_k \right]  \\ \nonumber & \quad\quad \cdot   p\left(x_k|\bbY_{k-1}\right) \rmd x_k\\ \nonumber
&= \int \left(x_k-\hat x_{k|k-1}\right) \left(h(x_k)-\hat y_{k|k-1} \right)^\T    p\left(x_k|\bbY_{k-1}\right) \rmd x_k \\
&= \cov \left[ x_k,h(x_{k}) | \bbY_{k-1} \right] .
%& = \E \left[\left. \left( x_k - \hat x_{k|k-1} \right) \left( h(x_k) - \hat y_{k|k-1} \right)^\T \right| \bbY_{k-1} \right]\\
%&= \int \left(x_k-\hat x_{k|k-1}\right) \left(h(x_k)-\hat y_{k|k-1} \right)^\T   \cdot \calN( x_k; \hat x_{k|k-1}, P_{k|k-1} ) \rmd x_k .
\end{align}
For two jointly Gaussian random vectors, the conditional distribution of one given another is also Gaussian, which is summarized in~\eqref{GRV_4} in Section~\ref{Linear-KF}.
It then follows from~\eqref{joint} that a Gaussian approximation can be constructed for $p\left(x_k|\bbY_k\right)$. Its mean and covariance can be expressed as
\begin{align}\label{GF_U_1}
\hat x_{k|k} &= \hat x_{k|k-1}  + \underbrace{P_{k|k-1}^{xy} \left(P_{k|k-1}^y\right)^{-1}}_{\rm Kalman \ gain} (y_k - \hat y_{k|k-1}),\\ \label{GF_U_2}
P_{k|k}^x &= P_{k|k-1}^x - P_{k|k-1}^{xy} \left(P_{k|k-1}^y\right)^{-1} \left(P_{k|k-1}^{xy} \right)^\T.
\end{align}

Putting together~\eqref{GF_P_1}-\eqref{GF_P_2}, \eqref{y_p}-\eqref{y_cov}
and~\eqref{xy_cov}-\eqref{GF_U_2}, we can establish a conceptual description of the KF technique, which is outlined in Fig.~\ref{Gaussian-filter}. Built in the Bayesian-Gaussian setting, it conducts state estimation through tracking the mean and covariance of a random state vector.
It is noteworthy that one needs to develop explicit expressions to enable the use of the above KF equations. The key that bridges the gap is to find the mean and covariance of a random vector passing through nonlinear functions. For linear dynamic systems, the development is straightforward, because, in the considered context the involved pdf's are strictly Gaussian and the linear transformation of the Gaussian state variables can be readily handled. The result is the standard KF to be shown in the next section. However, complications arise in the case of nonlinear systems. This issue has drawn significant interest from researchers. A wide range of ideas and methodologies have been developed, leading to a family of nonlinear KFs. The three most representative among them are EKF, UKF, and EnKF to be introduced following the review of the linear KF. 

\iffalse
\begin{figure*}[t] 
\makeatother
  \centering
  \subfigure[]{
    \includegraphics[trim={2cm 0.5cm 2.5cm 0.8cm},clip,width=0.23\linewidth]{System_State_Statistics_1.eps}\label{System_State_Statistics_1}}
  \subfigure[]{
    \includegraphics[trim={2cm 0.5cm 2.5cm 0.8cm},clip,width=0.23\linewidth]{System_State_Statistics_2.eps}\label{System_State_Statistics_2}}%
  \subfigure[]{
    \includegraphics[trim={2cm 0.5cm 2cm 0.9cm},clip,width=0.23\linewidth]{System_State_Statistics_3.eps}\label{System_State_Statistics_3}}%
    \subfigure[]{
    \includegraphics[trim={2cm 0.5cm 2.5cm 0.5cm},clip,width=0.23\linewidth]{System_Output_Statistics.eps}\label{System_Output_Statistics}}%
  \caption{The state and output pdf's of the considered linear dynamic system: (a) the Gaussian pdf of $x_0$; (b) the Gaussian pdf of $x_1$; (c) the Gaussian pdf of $x_2$; and (d) the Gaussian pdf's of $y_k$ for $k=\{ {\color{black}0},{\color{blue}1},{\color{red}2}\}$.}\label{Linear_System_Statistics}
\end{figure*}
\fi

\section{Standard Linear Kalman Filter}\label{Linear-KF}

In this section, KF for linear systems is reviewed briefly  to pave the way for discussion of nonlinear KFs.
Consider a linear time-invariant discrete-time system of the form
\begin{equation}\label{linear_sys}
\left\{
\begin{aligned}
x_{k+1} &= Fx_k+w_k,\\
y_k &= Hx_k+v_k,
\end{aligned}
\right.
\end{equation}
where: 1) $\{w_k\}$ and $\{v_k\}$ are zero-mean white Gaussian noise sequences with $w_k \sim \calN(0,Q)$ and $v_k \sim \calN(0,R)$, 2) $x_0$ is Gaussian with $x_0 \sim \calN(\bar x_0, P_0)$, and 3)
$x_0$, $\{w_k\}$ and $\{v_k\}$ are independent of each other. Note that, under these conditions, the Gaussian assumptions in Section~\ref{Bayesian-KF} will exactly hold for the linear system in~\eqref{linear_sys}.  %the equations in Section~\ref{Bayesian-KF} exactly hold for linear systems under these conditions. 

The standard KF for the linear dynamic system in~\eqref{linear_sys} can  be readily derived from the conceptual KF summarized in Fig.~\ref{Gaussian-filter}. Since the system is linear and under a Gaussian setting, $p\left( x_{k-1} | \bbY_{k-1}\right)$ and $p\left(x_k | \bbY_{k-1} \right)$ are strictly Gaussian according to the properties of Gaussian vectors. Specifically, $x_{k-1} | \bbY_{k-1} \sim \calN \left(\hat x_{k-1|k-1}, P_{k-1|k-1}^x \right)$ and $x_k | \bbY_{k-1} \sim \calN \left(\hat x_{k|k-1}, P_{k|k-1}^x \right) $. According to~\eqref{GF_P_1} and \eqref{GF_P_2}, the prediction is
\begin{align}\label{LKF_Prediction_1}
\hat x_{k|k-1} &= \E \left( Fx_{k-1} | \bbY_{k-1} \right)   = F\hat x_{k-1|k-1},\\ \label{LKF_Prediction_2}
\nonumber
P_{k|k-1}^x &= \cov\left( Fx_{k-1}|\bbY_{k-1}\right) +Q \\ & = FP_{k-1|k-1}^x F^\T+Q,
\end{align}

\begin{figure}[t] 
  \center
  \includegraphics[width=\linewidth]{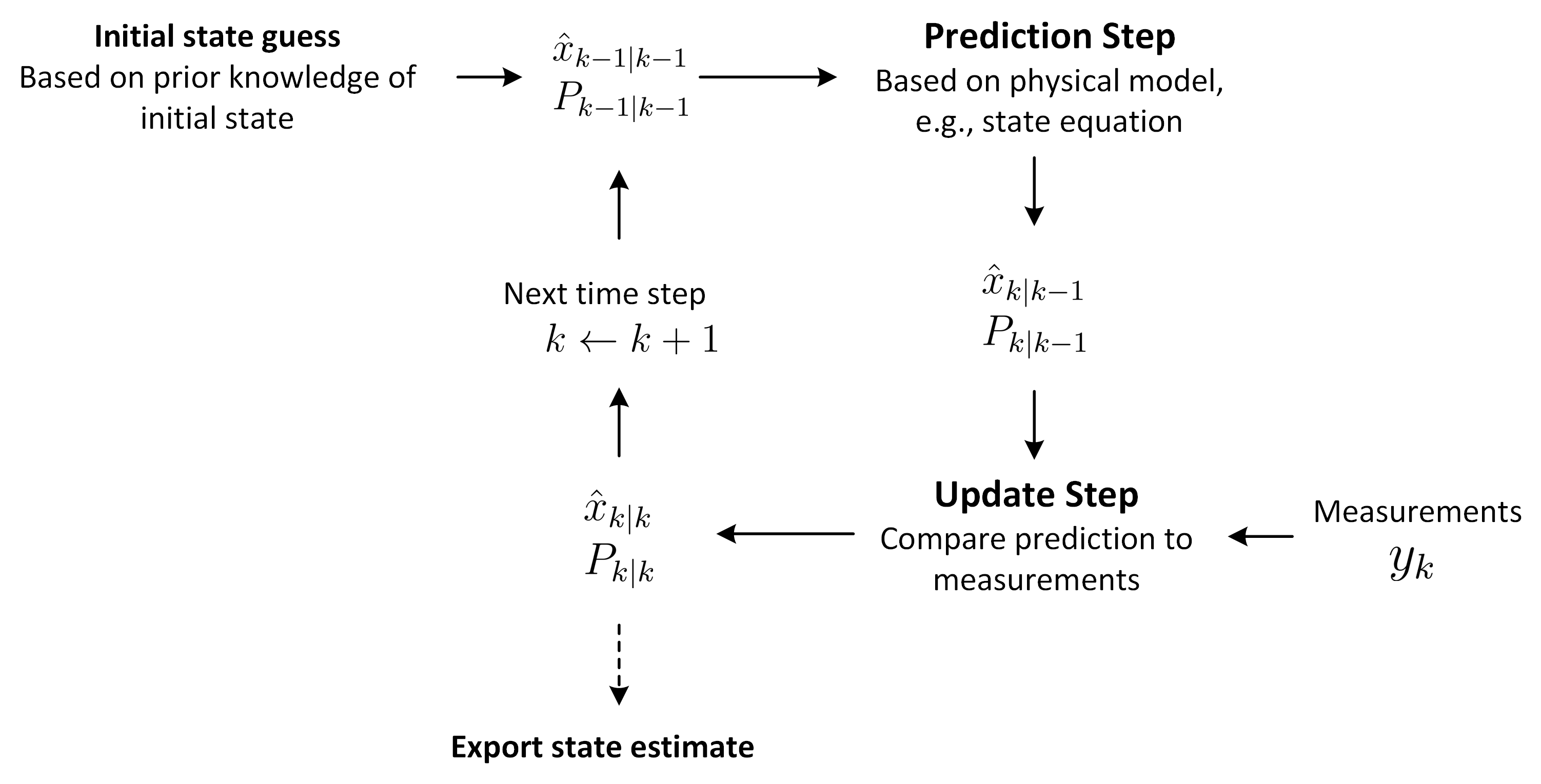}\\
  \caption[S1]{A schematic of the KF/EKF structure, modified from~\cite{KF:Wiki}. KF/EKF comprises two steps sequentially executed through time, prediction and update. For prediction, $x_k$ is predicted by using the data up to time $k-1$. The forecast is denoted as $\hat x_{k|k-1}$ and subject to uncertainty quantified by the prediction error covariance $P_{k|k-1}$. The update step occurs upon the arrival of the new measurement $y_k$. In this step, $y_k$ is leveraged to correct $\hat x_{k|k-1}$ and produce the updated estimate $\hat x_{k|k}$. Meanwhile, $P_{k|k-1}$ is updated to generate $P_{k|k}$ to quantify the uncertainty imposed on $\hat x_{k|k}$.}\label{KF-EKF-2}
\end{figure}

The update can be accomplished along the similar lines. Based on~\eqref{y_p}-\eqref{xy_cov}, we have $\hat y_{k|k-1} = H \hat x_{k|k-1}$, $P_{k|k-1}^y = H P_{k|k-1}H^\T+R$, and $P_{k|k-1}^{xy} = P_{k|k-1}H^\T$. Then, as indicated by~\eqref{GF_U_1}-\eqref{GF_U_2}, the updated state estimate is
\begin{align}\label{LKF_Update_1}\nonumber
\hat x_{k|k} &= \hat x_{k|k-1} + \underbrace{P_{k|k-1}^x H^\T \left( H P_{k|k-1}^x H^\T+R\right)^{-1}}_{\rm Kalman \ gain} \\ & \quad\quad \quad\quad \quad\quad \cdot \left(y_k-H \hat x_{k|k-1}\right),\\ \label{LKF_Update_2} \nonumber
P_{k|k}^x &= P_{k|k-1}^x - P_{k|k-1}^x H^\T \left( H P_{k|k-1}^x H^\T+R\right)^{-1} \\ & \quad\quad \quad\quad \quad\quad \cdot  H P_{k|k-1}^x.
\end{align}
Together,~\eqref{LKF_Prediction_1}-\eqref{LKF_Update_2} form the linear KF. Through the prediction-update procedure, it generates the state estimates and associated estimation error covariances recursively over time when the output measurement arrives. From the probabilistic perspective, $\hat x_{k|k}$ and $P_{k|k}$ together determine the Gaussian distribution of $x_k$ conditioned on $\bbY_k$. The covariance, quantifying how widely the random vector $x_k$ can potentially spread out, can be interpreted as a measure of the confidence on or uncertainty of the estimate. A schematic diagram of the KF is shown in Fig.~\ref{KF-EKF-2} (it can also be used to demonstrate EKF to be shown next).

Given that $(F,H)$ is detectable and $(F, Q^{1\over 2})$ stabilizable, $P_{k|k-1}^x$ converges to a fixed point, which is the solution to a discrete-time algebraic Riccati equation (DARE) 
\begin{align*}
X=FXF^\T-FXH^\T(HXH^\T+R)^{-1}HXF^\T+Q.
\end{align*}
This implies that KF can approach a steady state after a few time instants. With this idea, one can design a steady-state KF by solving  DARE offline to obtain the Kalman gain and then apply it to run  KF online, as detailed in~\cite{Anderson:1979}. Obviating the need for computing the gain and covariance at every time instant, the steady-state KF, though suboptimal, presents higher computational efficiency than the standard KF. %However, this desirable steady state will not arise for nonlinear KFs.

\section{Review of Nonlinear Kalman Filters}

In this section, an introductory overview of the major nonlinear KF techniques is offered, including the celebrated EKF and UKF in the field of control systems and the EnKF popular in the data assimilation community.

\subsection{Extended Kalman Filter}

EKF is arguably the most celebrated nonlinear state estimation technique, with numerous applications across a variety of engineering areas and beyond~\cite{Auger:TIE:2013}. It is based on the linearization of nonlinear functions around the most recent state estimate. When the state estimate $\hat x_{k-1|k-1}$ is generated, consider the first-order Taylor expansion of $f(x_{k-1})$ at this point:
\begin{align}\label{Taylor_f}
f(x_{k-1}) &\approx f(\hat x_{k-1|k-1}) + F_{k-1} \left(x_{k-1} - \hat x_{k-1|k-1} \right), \\ F_{k-1} &= \left. \frac{\partial f}{\partial x} \right|_{\hat x_{k-1|k-1}}.
\end{align}
For simplicity, let $p(x_{k-1}|\bbY_{k-1})$ be approximated by a distribution with mean $\hat x_{k-1|k-1}$ and covariance {\color{blue}$P_{k-1|k-1}^x$}. Then inserting~\eqref{Taylor_f} to~\eqref{GF_P_1}-\eqref{GF_P_2}, we can readily obtain the one-step-forward prediction
\begin{align}\label{EKF_P_1}
\hat x_{k|k-1} &= \E\left[ f \left(x_{k-1} | \bbY_{k-1} \right) \right] \approx  f\left( \hat x_{k-1|k-1} \right),\\\label{EKF_P_2}\nonumber
P_{k|k-1}^x&= \cov \left[ x_k | \bbY_{k-1} \right]+Q\\ & = F_{k-1} P_{k-1|k-1}^x F_{k-1}^\T+Q.
\end{align}

Looking into~\eqref{Taylor_f}, we find that the Taylor expansion approximates the nonlinear transformation of the random vector $x_k$ by an affine one. Proceeding on this approximation, we can easily estimate the mean and covariance of $f(x_{k-1})$ once provided the mean and covariance information of $x_{k-1}$ conditioned on $\bbY_{k-1}$. This, after being integrated with the effect of the noise $w_k$ on the prediction error covariance, establishes a prediction of $x_k$, as specified in~\eqref{EKF_P_1}-\eqref{EKF_P_2}. After $\hat x_{k|k-1}$ is produced, we can investigate the linearization of $h(x_k)$ around this new operating point in order to update the prediction. That is,
\begin{align}\label{Taylor_h}
h(x_k) &\approx h\left( \hat x_{k|k-1} \right) + H_k \left( x_k  - \hat x_{k|k-1} \right),  \\ H_k &= \left. \frac{\partial h}{\partial x} \right|_{\hat x_{k|k-1}}.
\end{align}
Similarly, we assume that $p(x_k |\bbY_{k-1})$ can be replaced by a  distribution with mean $\hat x_{k|k-1}$ and covariance $P_{k|k-1}^x$. Using~\eqref{Taylor_h},  the evaluation of~\eqref{y_p}-\eqref{xy_cov} yields $
\hat y_{k|k-1} \approx  h\left( \hat x_{k|k-1} \right)$, $
P_{k|k-1}^y \approx H_k P_{k|k-1}^x H_k^\T +R$, and
$P_{k|k-1}^{xy} \approx P_{k|k-1}^x H_k^\T$.

Here, the approximate mean and covariance information of $h(x_k)$ and $y_k$ are obtained through the linearization of $h(x_k)$ around $\hat x_{k|k-1}$. With the aid of the Gaussianity assumption in~\eqref{joint}, an updated estimate of $x_k$ is produced as follows:
\begin{align}\label{EKF_U_1}\nonumber
\hat x_{k|k} &= \hat x_{k|k-1} + P_{k|k-1}^x H_k^\T \left(H_k P_{k|k-1}^x H_k^\T+R \right)^{-1} \\ & \quad\quad\quad\quad\quad\quad\quad\quad \cdot \left[ y_k - h\left( \hat x_{k|k-1} \right) \right],\\ \label{EKF_U_2}\nonumber
P_{k|k}^x &= P_{k|k-1}^x -  P_{k|k-1}^x H_k^\T \left(H_k P_{k|k-1}^x H_k^\T+R \right)^{-1}\\  & \quad\quad\quad\quad\quad\quad\quad\quad \cdot  H_k P_{k|k-1}^x.
\end{align}
Then, EKF consists of~\eqref{EKF_P_1}-\eqref{EKF_P_2} for prediction and~\eqref{EKF_U_1}-\eqref{EKF_U_2} for update. When comparing it with the standard KF in~\eqref{LKF_Prediction_1}-\eqref{LKF_Update_2}, we can find that they share significant resemblance in structure, except that EKF introduces the linearization procedure to accommodate the system nonlinearities.

Since the 1960s, EKF has gained  wide use in the areas of aerospace, robotics, biomedical, mechanical, chemical, electrical and civil engineering, with great success in the real world witnessed. This is often ascribed to its relative easiness of design and execution. Another important reason is its good convergence from a theoretical viewpoint. In spite of the linearization-induced errors, EKF has provable asymptotic convergence under some conditions that can be satisfied by many practical systems~\cite{Boutayeb:TAC:1997,Krener:Springer:2003,Kluge:TAC:2010,Bonnabel:TAC:2015,Barrau:TAC:2017}. However, it also suffers from some shortcomings. The foremost one is the inadequacy of its first-order accuracy for highly nonlinear systems. In addition, the need for explicit derivative matrices not only renders EKF futile for discontinuous or other non-differentiable systems, but also pulls it away from convenient use in view of programming and debugging, especially when nonlinear functions of a complex structure are faced. This factor, together with the computational complexity at $O(n_x^3)$, limits the application of EKF to only low-dimensional systems.

Some modified EKFs have been introduced for improved accuracy or efficiency. In this regard, a natural extension is through the second-order Taylor expansion, which leads to the {\em second-order EKF} with more accurate estimation~\cite{Tanizaki:Springer:1996,Sarkka:CamPress:2013,Roth:Fusion:2011}. Another important variant, named {\em iterated EKF (IEKF)}, iteratively refines the state estimate around the current point at each time instant~\cite{Jazwinski:AP:1970,Bell:TAC:1993,Fang:CEP:2014}. Though requiring an increased computational cost, it can achieve higher estimation accuracy even in the presence of severe nonlinearities  in  systems.

%  --- it can be sensitive to the initial estimate, inaccurate owing to the approximation, and complex calculation caused by the linearization-induced derivative matrices, when the system functions $f(\cdot)$ and $h(\cdot)$ are highly nonlinear and/or high-dimensional. Overcoming such problems turns out to be a strong motivation for the research and development effort for more sophisticated state estimation techniques.

\subsection{Unscented Kalman Filter}

As the performance of EKF degrades for systems with strong nonlinearities, researchers have been seeking better ways to conduct nonlinear state estimation. In the 1990s,  UKF was invented~\cite{Julier:ACC:1995,Julier:AeroSense:1997}. Since then, it has been gaining significant popularity among researchers and practitioners. This technique is based on the so-called ``unscented transform (UT)'', which exploits the utility of deterministic sampling to track the mean and covariance information of a random variable passing through a nonlinear transformation~\cite{Julier:ProcIEEE:2004,Wan:Wiley:2001,Menegaz:TAC:2015}. The basic idea is to approximately represent a random variable by a set of sample points (sigma points) chosen deterministically to completely capture the mean and covariance. Then, projecting the sigma points through the nonlinear function concerned, one obtains a new set of sigma points and then use them to form the mean and covariance of the transformed variable for estimation. 

\iffalse
\begin{shadedbox}
{\Large Unscented Transform}
\vspace{3mm}
\fi

\begin{figure}[t]
  \center
  \includegraphics[width=\linewidth]{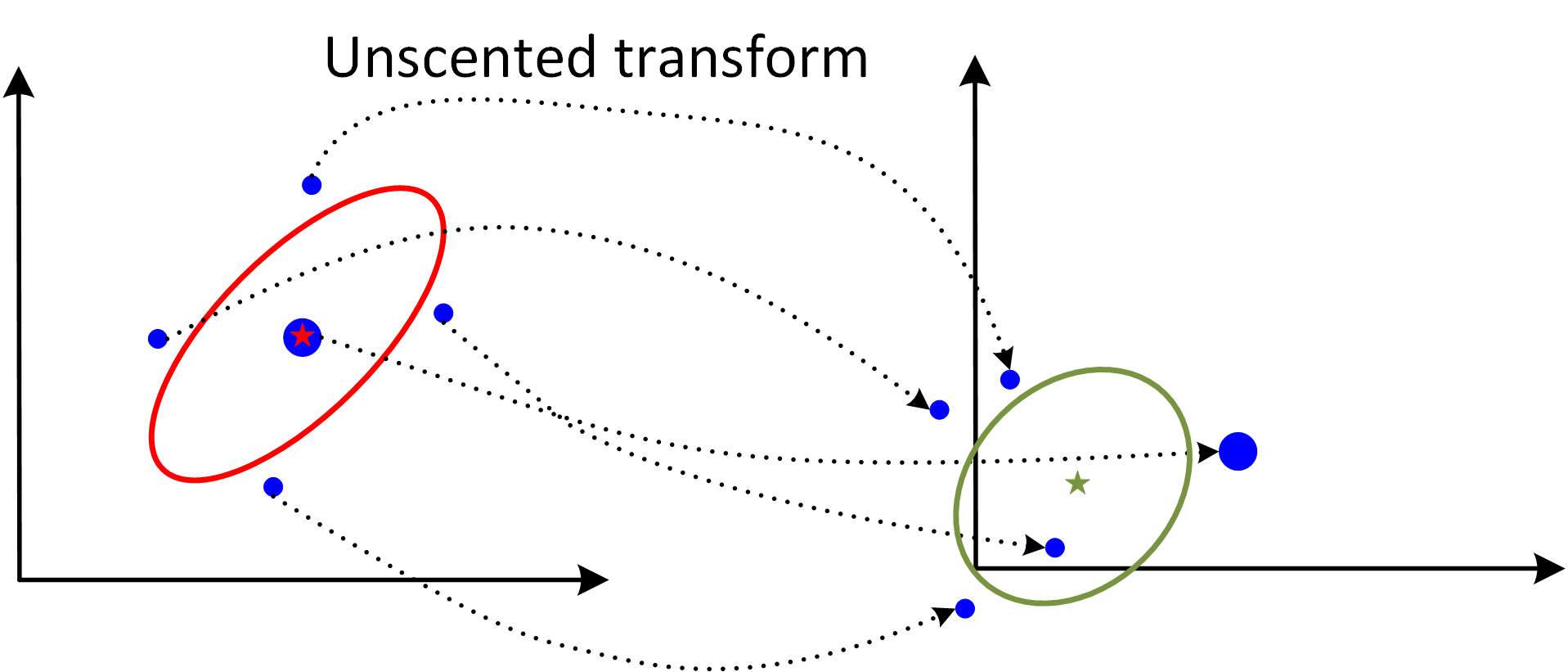}\\
  \caption{A schematic sketch of the UT procedure, adapted from~\cite{Julier:ProcIEEE:2004}. A set of sigma points (blue dots) are generated first according to the initial mean (red five-pointed star) and covariance (red ellipse) (left) and projected through the nonlinear function to generate a set of new sigma points (right). The new sigma points are then used to calculate the new mean (green star) and covariance (green ellipse).}\label{UT-Schematic}
\end{figure}

To explain how UT tracks the statistics of a nonlinearly transformed random variable, we consider a random variable $x\in \bbR^n$ and a nonlinear function $z = g(x)$. It is assumed that the mean and covariance of $x$ are $\bar x$ and $P_x$, respectively. The UT proceeds as follows~\cite{Julier:ProcIEEE:2004,Wan:Wiley:2001}.
First, a set of sigma points $\{x^i, i=0,1,\cdots,2n\}$ are chosen deterministically:
\begin{align}
x^0 & = \bar x,\\
x^i &= \bar x+\sqrt{n+\lambda}\left[ \sqrt {P_x} \right]_i, \ i=1,2,\cdots,n,\\
x^{i+n} &= \bar x-\sqrt{n+\lambda}\left[ \sqrt {P_x} \right]_i, \ i=1,2,\cdots,n,
\end{align}
where $[\cdot]_i$ represents the $i$-th column of the matrix and the matrix square root is defined by $\sqrt{P} \sqrt{P}^\T = P$ achievable through the Cholesky decomposition. The sigma points spread across the space around $\bar x$. The width of spread is dependent on the covariance $P$ and the scaling parameter $\lambda$, where $\lambda=\alpha^2 (n+\kappa) - n$. Typically, $\alpha$ is a small positive value (e.g., $10^{-3}$), and $\kappa$ is usually set to $0$ or $3-n$~\cite{Julier:ACC:1995}.
Then the sigma points are propagated through the nonlinear function $g(\cdot)$ to generate the sigma points for the transformed variable $z$, i.e.,
\begin{align*}
z^i = g\left(x^i\right), \ i=0,1,\cdots,2n.
\end{align*}
The mean and covariance of $z$ are estimated as
\begin{align}
\bar z &= \E\left[g(x)\right]\approx \sum_{i=0}^{2n} W_i^{(m)} z^i,\\ \nonumber
P_z &= \E\left[\left(g(x)-\bar z\right)\left(g(x)-\bar z\right)^\T\right]\\ & \approx \sum_{i=0}^{2n} W_i^{(c)} \left(z^i - \bar z \right) \left(z^i - \bar z \right)^\T,
\end{align}
where the weights are
\begin{align}\label{UT_Weight_1}
W_0^{(m)} &= {\lambda\over n+\lambda},\\ \label{UT_Weight_2}
W_0^{(c)} &= \frac{\lambda}{n+\lambda} + (1-\alpha^2+\beta),\\ \label{UT_Weight_3}
W_i^{(m)} &=W_i^{(c)} = \frac{1}{2(n+\lambda)}, \ i = 1,2,\cdots, 2n.
\end{align}
The parameter $\beta$ in~\eqref{UT_Weight_2} can be used to include prior information on the distribution of $x$. When $x$ is Gaussian, $\beta=2$ is optimal. The UT procedure is schematically shown in Fig.~\ref{UT-Schematic}.

\iffalse

\end{shadedbox}
\fi

To develop UKF, it is necessary to apply UT at both prediction and update steps, which involve nonlinear state transformations based on $f$ and $h$, respectively. For prediction, suppose that the mean and covariance of $x_{k-1}$, $\hat x_{k-1|k-1}$ and $P_{k-1|k-1}^x$, are given. To begin with, the sigma points for $x_{k-1}$ are generated:
\begin{align}\label{UKF_P_Simga_1}
\hat x_{k-1|k-1}^{(0)} & = \hat x_{k-1|k-1},\\\label{UKF_P_Simga_2}\nonumber
\hat x_{k-1|k-1}^{(i)} &= \hat x_{k-1|k-1}+\sqrt{n_x+\lambda}\left[ \sqrt {P_{k-1|k-1}^x} \right]_i, \\& \quad\quad\quad\quad\quad\quad \quad\quad\quad  i=1,2,\cdots,n_x,\\\label{UKF_P_Simga_3}
\hat x_{k-1|k-1}^{(i+n_x)} &= \hat x_{k-1|k-1} -\sqrt{n_x+\lambda}\left[ \sqrt {P_{k-1|k-1}^x}\nonumber \right]_i, \\& \quad\quad\quad\quad\quad\quad \quad\quad\quad  \ i=1,2,\cdots,n_x.
\end{align}
Then, they are propagated forward through the nonlinear function $f(\cdot)$, that is,
\begin{align}\label{UKF_P_1}
\hat x_{k|k-1}^{-(i)} = f\left(\hat x_{k-1|k-1}^{(i)}\right), \ i=0,1,\cdots,2n_x.
\end{align}
These new sigma points are considered capable of capturing the mean and covariance of $f(x_{k-1})$. Using them, the prediction of $x_k$ can be achieved as follows:
\begin{align}\label{UKF_P_2}
&\hat x_{k|k-1} = \E\left[ f \left(x_{k-1}\right) | \bbY_{k-1}  \right] =  \sum_{i=0}^{2n_x} W_i^{(m)} \hat x_{k|k-1}^{-(i)},\\\label{UKF_P_3}\nonumber
&P_{k|k-1}^x = \cov\left[ f \left(x_{k-1}\right) | \bbY_{k-1}  \right]+Q\\ \nonumber & =\sum_{i=0}^{2n_x} W_i^{(c)} \left( \hat x_{k|k-1}^{-(i)} - \hat x_{k|k-1} \right)\left( \hat x_{k|k-1}^{-(i)} - \hat x_{k|k-1} \right)^\T \\ & \quad\quad\quad\quad +Q.
\end{align}
%where the weights $W_i^{(m)}$ and $W_i^{(c)}$ are defined in~\eqref{UT_Weight_1}-\eqref{UT_Weight_3}.

By analogy, the sigma points for $x_k$ need to be generated first in order to perform the update, which are
\begin{align}\label{UKF_U_Sigma_1}
\hat x_{k|k-1}^{+(0)} & = \hat x_{k|k-1},\\\label{UKF_U_Sigma_2}\nonumber
\hat x_{k|k-1}^{+(i)} &= \hat x_{k|k-1}+\sqrt{n_x+\lambda}\left[ \sqrt {P_{k|k-1}^x} \right]_i, \\& \quad\quad\quad\quad\quad\quad \quad\quad\quad  i=1,2,\cdots,n_x,\\\label{UKF_U_Sigma_3}\nonumber
\hat x_{k|k-1}^{+(i+n_x)} &= \hat x_{k|k-1} -\sqrt{n_x+\lambda}\left[ \sqrt {P_{k|k-1}^x} \right]_i, \\& \quad\quad\quad\quad\quad\quad \quad\quad\quad  i=1,2,\cdots,n_x.
\end{align}
Propagating them through $h(\cdot)$, we can obtain the sigma points for $h(x_k)$, given by
\begin{align}\label{UKF_U_1}
\hat y_{k|k-1}^{(i)} &= h\left(\hat x_{k|k-1}^{+(i)}\right), \ i = 0,1,\cdots,2n_x.
\end{align}
The predicted mean and covariance of $y_k$ and the predicted cross-covariance between $x_k$ and $y_k$ are as follows:
\begin{align}\label{UKF_U_2}
&\hat y_{k|k-1} = \E \left[ y_k | \bbY_{k-1} \right]=\sum_{i=0}^{2n_x} W_i^{(m)} \hat y_{k|k-1}^{(i)},\\\label{UKF_U_3}\nonumber
&P_{k|k-1}^y = \cov\left[ h(x_k) | \bbY_{k-1}  \right]+R\\&=\sum_{i=0}^{2n_x} W_i^{(c)} \left( \hat y_{k|k-1}^{(i)} - \hat y_{k|k-1} \right)\left( \hat y_{k|k-1}^{(i)} - \hat y_{k|k-1} \right)^\T+R,\\\label{UKF_U_4}
&P_{k|k-1}^{xy}= \cov\left[ x_k, h(x_k) | \bbY_{k-1}  \right]\\& =\sum_{i=0}^{2n_x} W_i^{(c)} \left( \hat x_{k|k-1}^{+(i)} - \hat x_{k|k-1} \right)\left( \hat y_{k|k-1}^{(i)} - \hat y_{k|k-1} \right)^\T,
\end{align}
With the above quantities becoming available, the Gaussian update in~\eqref{GF_U_1}-\eqref{GF_U_2} can be leveraged to enable the projection from the predicted estimate $\hat x_{k|k-1}$ to the updated estimate $\hat x_{k|k}$.

\begin{figure}[t]
  \center
  \includegraphics[width=\linewidth]{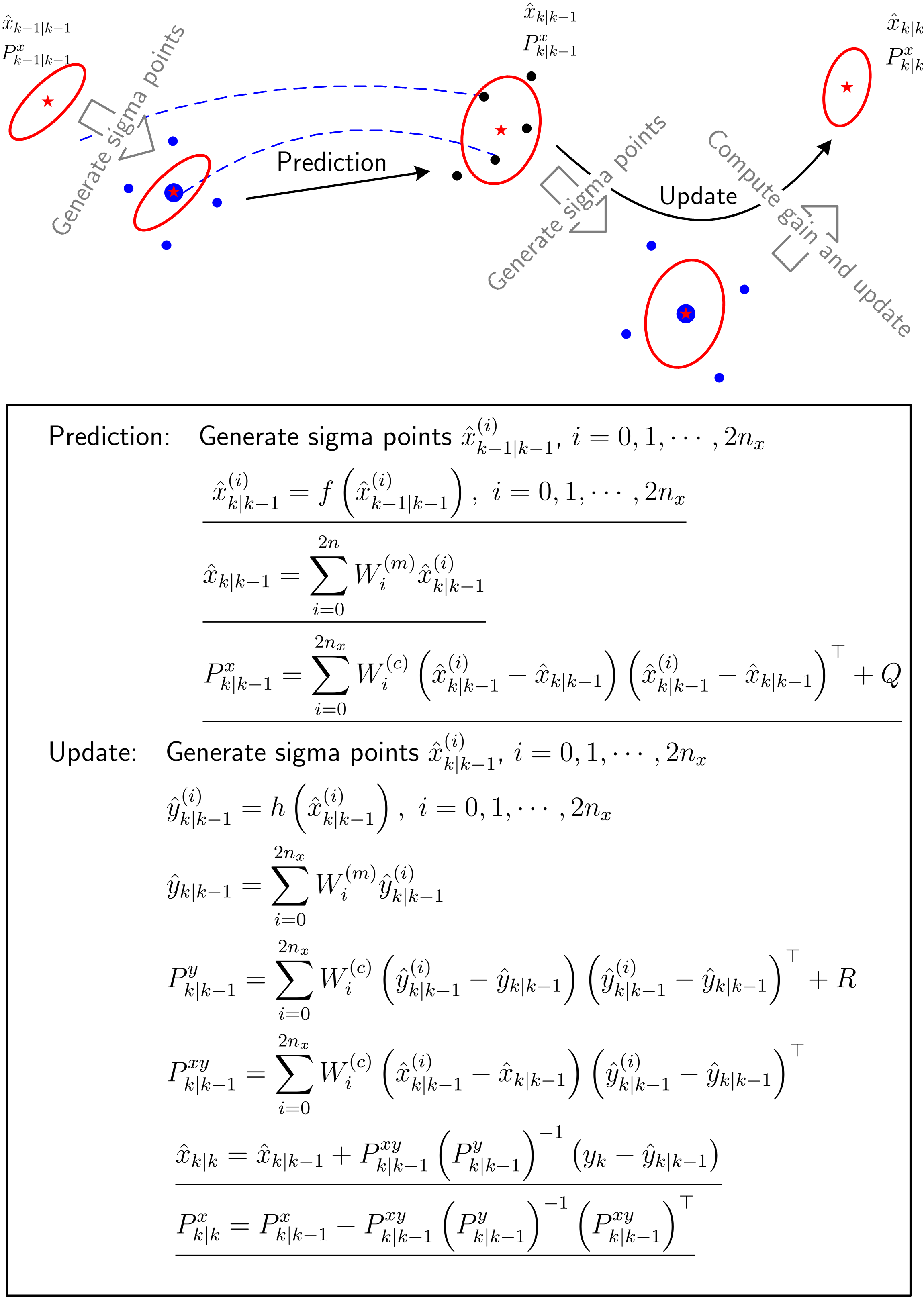}\\
  \caption{A schematic of UKF. Following the prediction-update procedure, UKF tracks the mean and covariance of state $x_k$ using sigma points chosen deterministically. A state estimate is graphically denoted by a red five-pointed star mean surrounded by a covariance ellipse, and the sigma points are colored in blue dots. }%For prediction, sigma points are generated for $\hat x_{k-1|k-1}$ and projected.}
  \label{Unscented-KF}
\vspace{-6mm}
\end{figure}

Summarizing the above equations leads to UKF sketched in Fig.~\ref{Unscented-KF}. Compared with EKF, UKF incurs a computational cost of the same order $O(n_x^3)$ but offers second-order accuracy~\cite{Julier:ProcIEEE:2004}, implying an overall smaller estimation error. In addition, its operations are derivative-free, exempt from the burdensome calculation of the Jacobian matrices in EKF. This will not only bring convenience to practical implementation but also imply its applicability to discontinuous undifferentiable nonlinear transformations. Yet, it is noteworthy that, with a complexity of $O(n_x^3)$ and operations of $2n_x+1$ sigma points, UKF faces substantial computational expenses when the system is high-dimensional with a large $n_x$, thus unsuitable for this kind of estimation problems.

Owing to its merits, UKF has seen a growing momentum of research since its advent. A large body of work is devoted to the development of modified versions. In this respect, {\em square-root UKF} (SR-UKF) directly propagates a square root matrix, which enjoys better numerical stability than squaring the propagated covariance matrices~\cite{Merwe:ICASSP:2001}; iterative refinement of the state estimate can also be adopted to enhance UKF as in IEKF , leading to {\em iterated UKF} (IUKF). The performance of UKF can be improved by selecting the sigma points in different ways. While the standard UKF employs symmetrically distributed $2n_x+1$ sigma points, asymmetric point sets or sets with a larger number of points may bring better accuracy~\cite{Julier:ACC:2003,Li:ISCIT:2007,Date:MTNS:2010,Charalampidis:CTA:2011}. Another interesting question is the determination of the optimal scaling parameter $\kappa$, which is investigated in~\cite{Dunik:TAC:2012}. UKF can be generalized to the so-called {\em sigma-point Kalman filtering} (SPKF), which refers to the class of filters that uses deterministic sampling points to determine the mean and covariance of a random vector through nonlinear transformation~\cite{Crassidis:TAES:2006,Merwe:Thesis:2004}. Other SPKF techniques include the central-difference Kalman filter (CDKF) and Gauss-Hermite filter (GHKF), which perform sigma-point-based filtering and can also be interpreted from the perspective of Gaussian-quadrature-based filtering~\cite{Ito:TAC:2000} (GHKF will receive further discussion in Section~\ref{Other-Problems}).

\subsection{Ensemble Kalman Filter}

Since its early development in~\cite{Evensen:JGRC:1994,Evensen:MWR:1996,Houtekamer:MWR:1998}, ensemble Kalman filter (EnKF) has established a strong presence in the field of state estimation for large-scale nonlinear systems. Its design is built on an integration of KF with the Monte Carlo method, which is a prominent statistical method concerning
simulation-based approximation of probability distributions using samples directly drawn from certain distributions. Basically, EnKF maintains an ensemble representing the conditional distribution of a random state vector given the measurement set. The state estimate is generated from the sample mean and covariance of the ensemble. In view of the sample-based approximation of probability distributions, EnKF shares similarity with UKF; however, the latter employs deterministic sampling while EnKF adopts non-deterministic sampling.

%While the GF only tracks the mean and covariance rather than an ensemble, it shares much similarity with the EnKF in terms of the Gaussian update adopted to correct the prediction.

Suppose that there is
an ensemble of samples, $ \hat x_{k-1|k-1}^{(i)}$ for $i=1,2,\cdots, N_s $, drawn from $p(x_{k-1} | \bbY_{k-1})$ to approximately represent this pdf.
Next, let an ensemble of samples, $\{w_{k-1}^{(i)} \}$ for $i=1,2,\cdots,N_s$, be drawn independently and identically from the
Gaussian distribution $\calN(0,Q)$ in order to
account for the process noise $w_{k-1}$ at time $k-1$.
%We assume that
%\begin{align*}
%p(x_k | x_{k-1}) = \frac{1}{N_s} \sum_{i=1}^{N_s} \delta \left( x_k - f(x_{k-1}) - w_{k-1}^i \right).
%\end{align*}
Then, $\hat x_{k-1|k-1}^{(i)} $ can hence be
projected to generate {\em a priori}
ensemble $\left\{ \hat
x_{k|k-1}^{(i)} \right\}$ that represents $p(x_k | \bbY_{k-1})$ as follows:
\begin{align}\label{EnKF_Prediction_1}
\hat x_{k|k-1}^{(i)} = f\left(\hat x_{k-1|k-1}^{(i)} \right) + w_{k-1}^{(i)}, \ i=1,2,\cdots, N_s
\end{align}
The sample mean and covariance of this ensemble can be calculated as:
\begin{align}\label{EnKF_Prediction_2}
&\hat x_{k|k-1} = \E \left(x_k | \bbY_{k-1} \right) \approx \frac{1}{N_s} \sum_{i=1}^{N_s} \hat x_{k|k-1}^{(i)},\\ \label{EnKF_Prediction_3}\nonumber
&P_{k|k-1}^x = \cov \left(x_k | \bbY_{k-1} \right)\\& \approx \frac{1}{N_s-1} \sum_{i=1}^{N_s} \left( \hat x_{k|k-1}^{(i)} -\hat x_{k|k-1} \right)\left( \hat x_{k|k-1}^{(i)} -\hat x_{k|k-1} \right)^\T , %+ Q_{k-1},
\end{align}
which form the prediction formulae.

The update step begins with the construction of the ensemble for $p\left(y_k | \bbY_{k-1} \right)$ by means of
\begin{align} \label{EnKF_Update_1}
    \hat y_{k|k-1}^{(i)} = h \left( \hat x_{k|k-1}^{(i)} \right)+v_k^{(i)}, \ i=1,2,\cdots, N_s
\end{align}
where $v_k^i $ is generated as per the Gaussian
distribution $\calN(0,R)$ to delineate the measurement noise $v_k$. The sample mean of this ensemble is
\begin{align}\label{EnKF_Update_2}
\hat y_{k|k-1} = \frac{1}{N_s} \sum_{i=1}^{N_s} \hat y_{k|k-1}^{(i)},
\end{align}
with the associated sample covariance
\begin{align} \label{EnKF_Update_3}\nonumber
P_{k|k-1}^y &= \frac{1}{N_s-1} \sum_{i=1}^{N_s} \left( \hat y_{k|k-1}^{(i)} - \hat y_{k|k-1}\right) \\ & \quad\quad\quad\quad \cdot \left( \hat y_{k|k-1}^{(i)} - \hat y_{k|k-1}\right)^\T.
\end{align}\label{EnKF_Update_4}
The cross-covariance between $x_k$ and $y_k$ given $\bbY_{k-1}$ is
\begin{align} \label{EnKF_Update_5}\nonumber
P_{k|k-1}^{xy} &= \frac{1}{N_s-1} \sum_{i=1}^{N_s}
\left( \hat x_{k|k-1}^{(i)} -   \hat x_{k|k-1}  \right)  \\ & \quad\quad\quad\quad \cdot \left( \hat y_{k|k-1}^{(i)} -   \hat y_{k|k-1} \right)^\T.
\end{align}
Once it arrives, the latest measurement $y_k$ can be applied to update each member of {\em a priori} ensemble in the way defined by~\eqref{GF_U_1}, i.e.,
\begin{align} \label{EnKF_Update_6}\nonumber
\hat x_{k|k}^{(i)} &= \hat x_{k|k-1}^{(i)} + P_{k|k-1}^{xy} \left(P_{k|k-1}^{y}\right)^{-1}
\left( y_k - \hat y_{k|k-1}^{(i)} \right), \\& \quad\quad\quad\quad\quad\quad \quad\quad\quad  i=1,2,\cdots,N_s.
\end{align}
This {\em a posteriori} ensemble $\left\{\hat x_{k|k}^{(i)}\right\}$ can be regarded as an approximate representation of
$p(x_k | \bbY_k)$. Then,
the updated estimation of the mean and covariance of $x_k$ can be achieved by
\begin{align}\label{EnKF_Update_7}
\hat x_{k|k} &= \frac{1}{N_s} \sum_{i=1}^{N_s} \hat x_{k|k}^i,\\ \label{EnKF_Update_8}
P_{k|k}^x &= \frac{1}{N_s-1} \sum_{i=1}^{N_s}
\left(  \hat x_{k|k}^{(i)} -  \hat x_{k|k} \right) \left(  \hat x_{k|k}^{(i)} -  \hat x_{k|k} \right)^\T.
\end{align}

The above ensemble-based prediction and update will be repeated recursively, forming EnKF. Note that the computation of estimation error covariance in~\eqref{EnKF_Prediction_3} and~\eqref{EnKF_Update_8} can be skipped if a user has no interest in learning about the estimation accuracy. This can further cut down EnKF's storage and computational cost.

\begin{figure}[t]
  \centering
  \includegraphics[width=\linewidth]{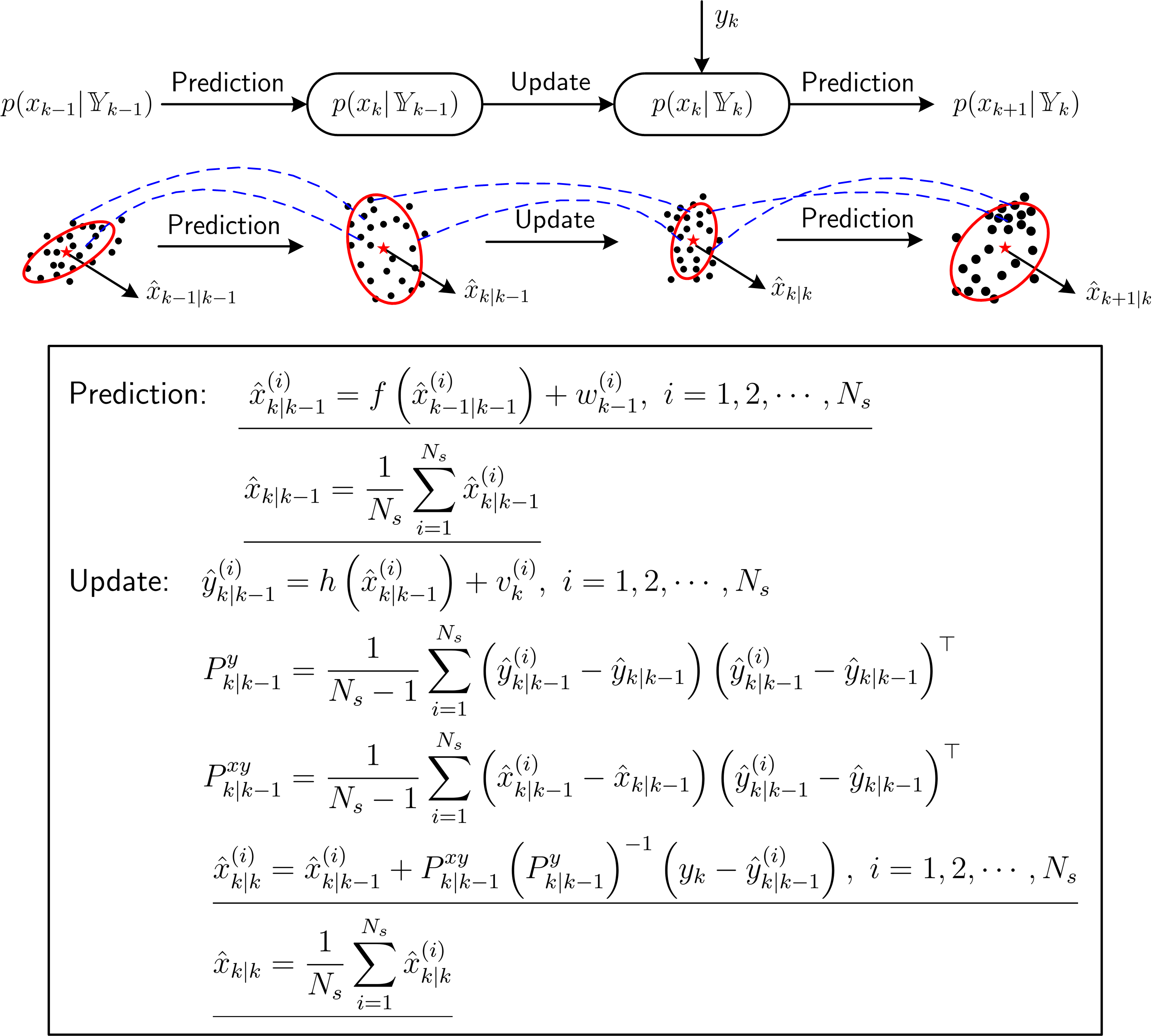}\\
  \caption{A schematic of EnKF. EnKF maintains an ensemble of sample points for the state vector $x_k$. It propagates and updates the ensemble to track the distribution of $x_k$. The state estimation is conducted by calculating the sample mean (red five-pointed-star) and covariance (red ellipse) of the ensemble.}\label{Ensemble-Kalman-filter}
\end{figure}

EnKF is illustrated schematically in
Fig.~\ref{Ensemble-Kalman-filter}. It features direct operation on the
ensembles as a Monte Carlo-based extension of KF. Essentially, it
represents the pdf of a state vector by using an ensemble of samples, propagates the ensemble members
and makes estimation by computing the mean and covariance of
the ensemble at each time instant.  Its complexity is $O(n_y^3+n_y^2 N_s+n_y N_s^2 +n_x N_s^2)$ ($n_x \gg n_y$ and $nx\gg N_s$ for high-dimensional systems)~\cite{Mandel:TechReport:2006}, which contrasts with $O(n_x^3)$ of EKF and UKF. This, along with the derivative-free computation and freedom from covariance matrix propagation, makes EnKF computationally efficient and appealing to be the method of choice for high-dimensional nonlinear systems.
An additional contributing factor in this respect is that the ensemble-based computational structure places it in an advantageous position for parallel implementation~\cite{Lakshmivarahan:CSM:2009}.
It has been
reported that convergence of the EnKF can be fast even with a
reasonably small ensemble
size~\cite{Evensen:Springer:2009,Mandel:AOM:2011}. In particular, its convergence to KF
in the limit for large ensemble size and Gaussian state probability
distributions is proven in~\cite{Mandel:AOM:2011}.

{\color{blue}
\section{Application to Speed Sensorless Induction Motors}

This section presents a case study of applying EKF, UKF and EnKF to state estimation for speed sensorless induction motors. Induction motors are used as an enabling component for numerous industrial systems, e.g., manufacturing machines, belt conveyors, cranes,
lifts, compressors, trolleys, electric vehicles, pumps, and fans~\cite{Montanari:TCST:2007}. In an induction motor, electromagnetic induction from the magnetic field of the stator winding is used to generate the electric current that drives the rotor to produce torque. This dynamic process must be delicately controlled to ensure accurate and responsive operations. Hence, control design for this application was researched extensively during the past decades, e.g.,~\cite{Bowes:TIE:2004,Marino:AUTO:2004,Montanari:TCST:2007}. Recent years have seen a growing interest in speed sensorless induction motors, which have no sensors to measure the rotor speed to reduce costs and increase reliability. However, the absence of the rotor speed makes control design more challenging. To address this challenge, state estimation is exploited to recover the speed and other unknown variables. It is also noted that an induction motor as a multivariable and highly nonlinear system makes a valuable benchmark for evaluating different state estimation approaches~\cite{Wang:AIM:2016,Marino:AUTO:2004}.

\begin{figure*}[t] 
\makeatother
  \centering
  \subfigure[]{
    \includegraphics[trim={.5cm 0.1cm 1cm 0.6cm},clip,width=0.32\linewidth]{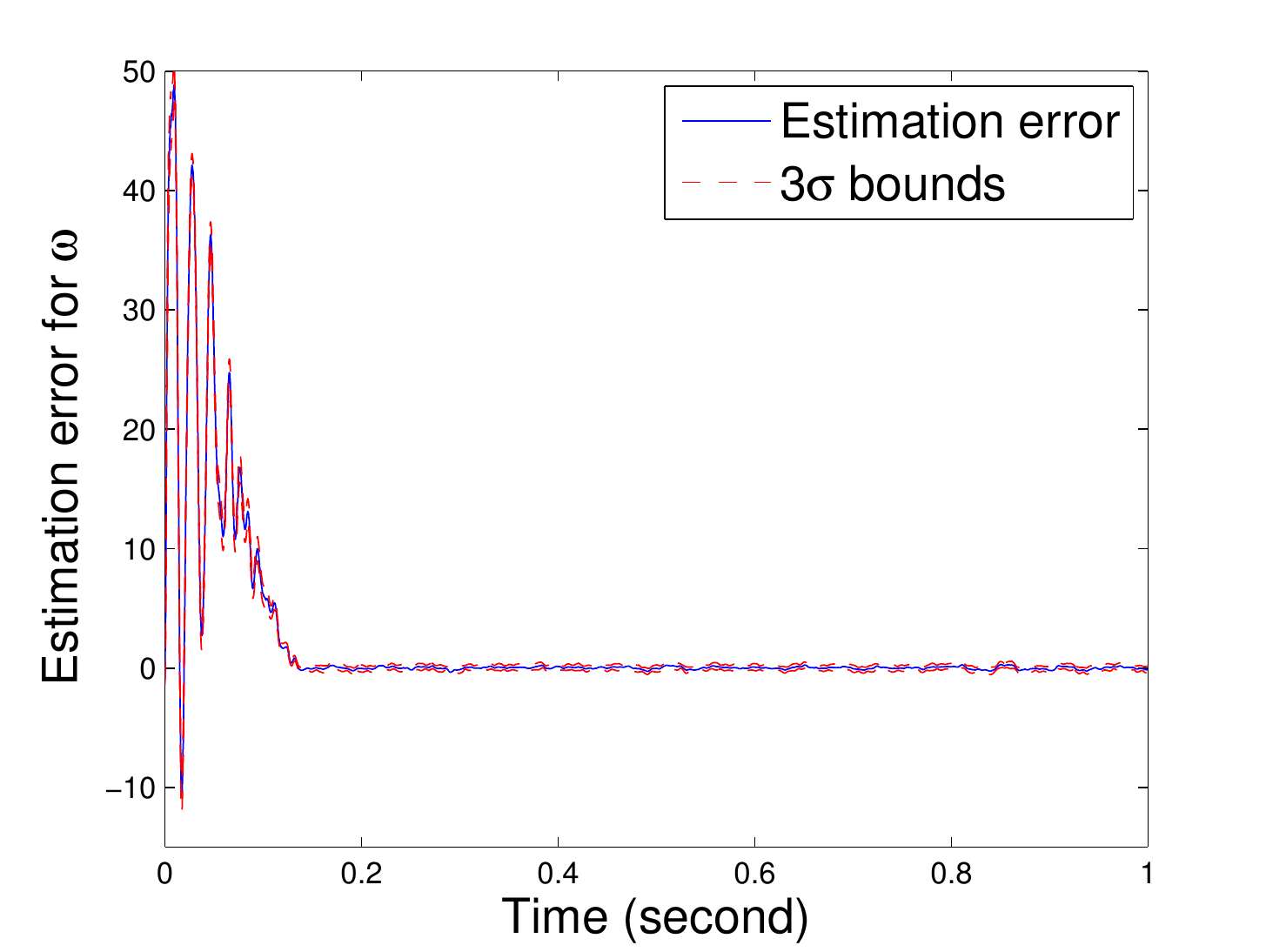}\label{EKF-estimation-error}}
  \subfigure[]{
    \includegraphics[trim={.5cm 0.1cm 1cm 0.6cm},clip,width=0.32\linewidth]{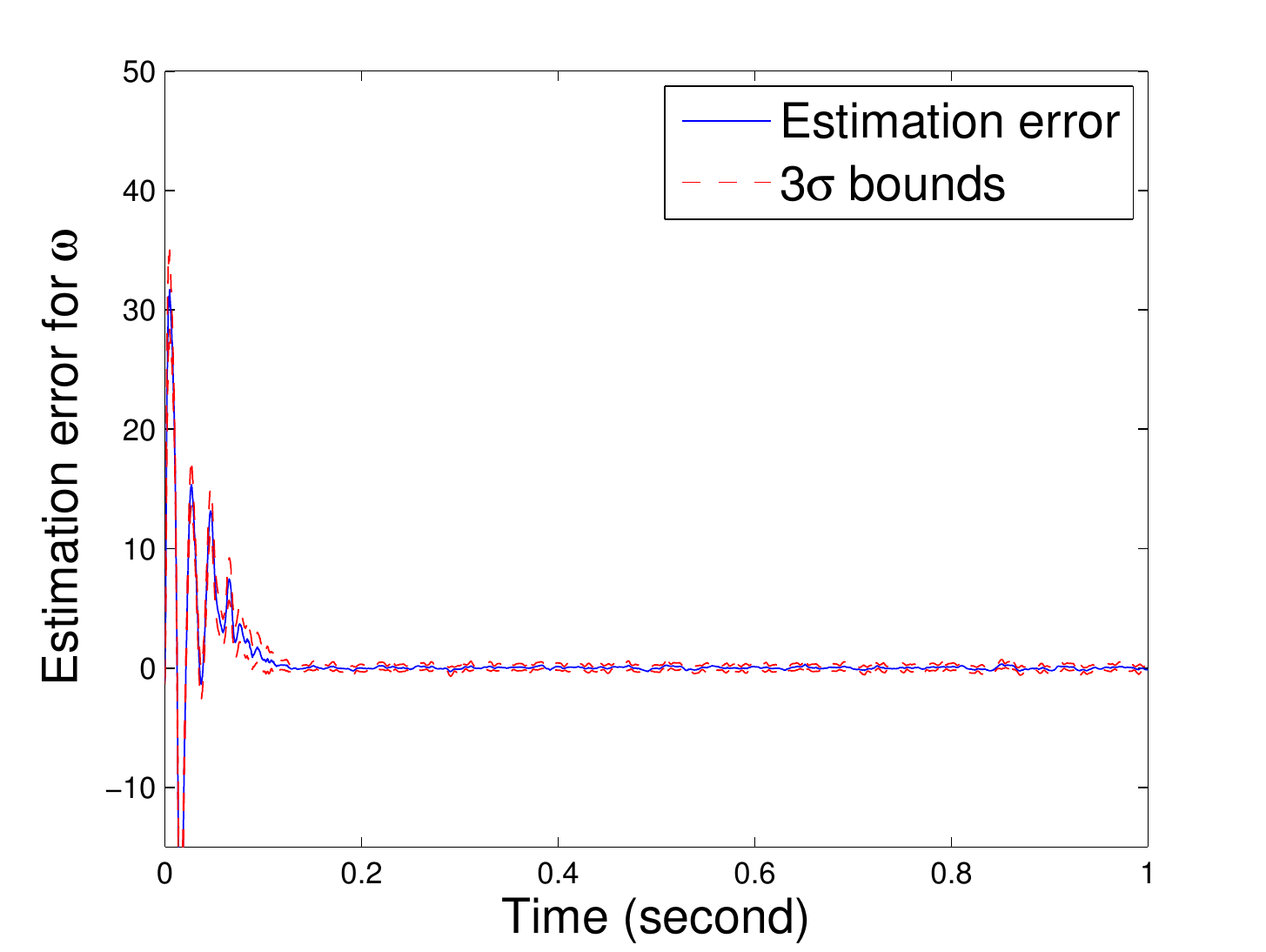}\label{UKF-estimation-error}}%
  \subfigure[]{
    \includegraphics[trim={.5cm 0.1cm 1cm 0.6cm},clip,width=0.32\linewidth]{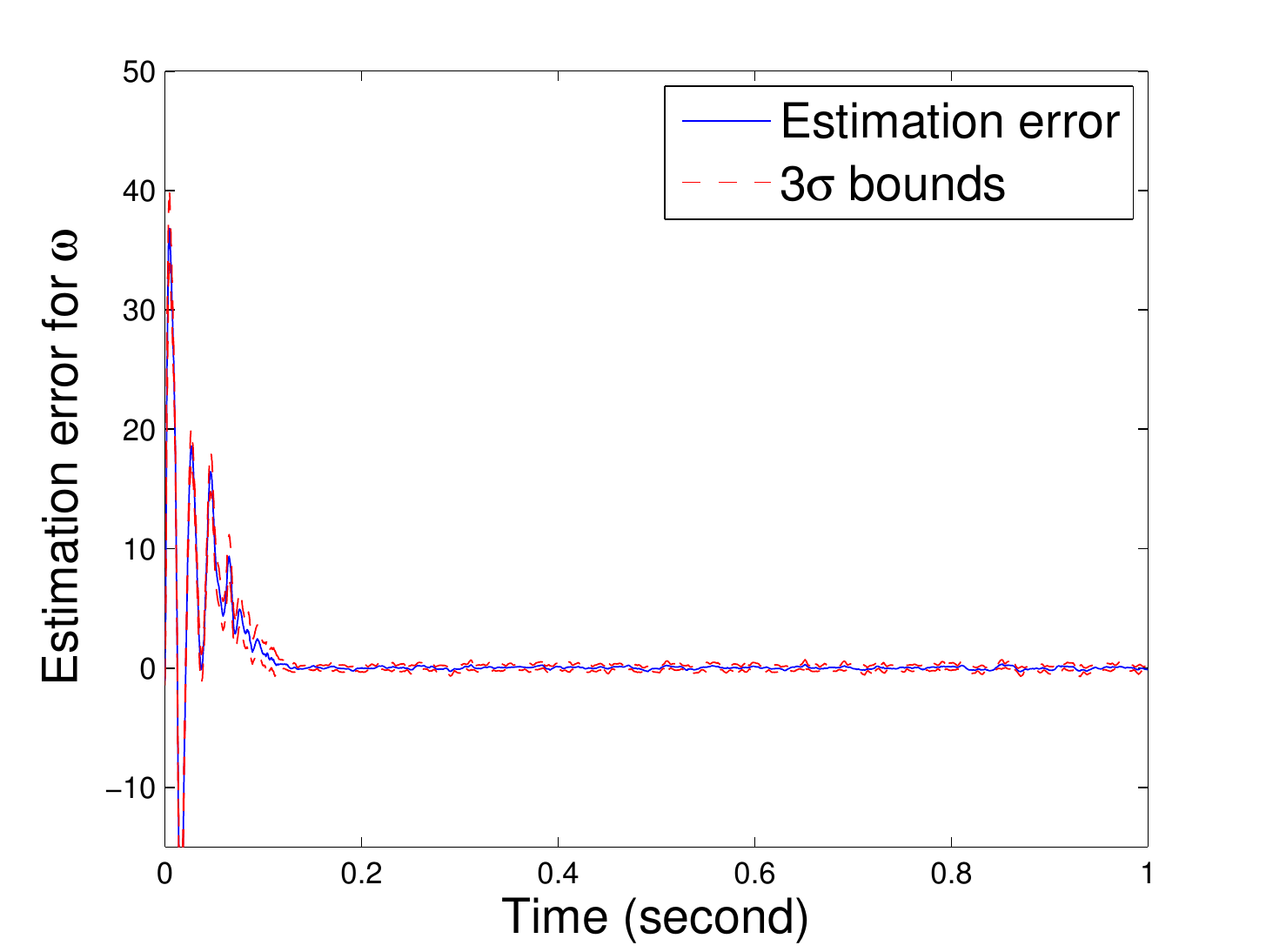}\label{EnKF-estimation-error}}
  \caption{\color{blue}Estimation error for $\omega$: (a) EKF; (b) UKF; (c) EnKF with $N_s=100$. }
  \label{Estimation-error}
\end{figure*}

\begin{table*}\color{blue}
\begin{center}
\caption{\color{blue}Average estimation errors for EKF, UKF, and EnKF.}\label{EKF-UKF-EnKF-Error}
\begin{tabular}{c|c|c|c|c|c|c}\hline 
\multirow{2}*{Filter } &\multirow{2}*{ EKF} & \multirow{2}*{UKF} & \multicolumn{4}{|c}{EnKF}\\ \cline{4-7}
&&&$N_s=40$&$N_s=100$&$N_s=200$&$N_s=400$\\ \hline
Average estimation error &           682.63 & 321.05 &           347.24 &   329.70 &  323.96 & 320.30\\ \hline
\end{tabular}
\end{center}
\end{table*}

The induction motor model in a stationary two-phase reference frame can be written as
\begin{align*}
\dot i_{ds} &= -\gamma i_{ds} + \alpha \beta \psi_{dr} +\beta \psi_{qr} \omega+ u_{ds}/\sigma,\\
\dot i_{qs} &= -\gamma i_{qs} - \beta \psi_{dr}\omega +\alpha \beta \psi_{qr}+ u_{qs}/\sigma,\\
\dot \psi_{dr} &= \alpha L_m i_{ds} - \alpha \psi_{dr} - \psi_{qr}\omega,\\
\dot \psi_{qr} &= \alpha L_m i_{qs} + \psi_{dr} \omega -\alpha \psi_{qr},\\
\dot \omega &= \frac{\mu}{J}\left( - i_{ds} \psi_{qr} + \psi_{dr} i_{qs} \right) - \frac{T_L}{J},\\
y & = \left[ \begin{matrix} i_{ds} \cr i_{qs} \end{matrix} \right],
\end{align*}
where $(\psi_{dr},\psi_{qr})$ is the rotor fluxes, $ (i_{ds}, i_{qs})$ is the stator currents, and $(u_{ds}, u_{qs})$ is the stator voltages, all defined in a stationary $d$-$q$ frame. In addition, $\omega$ is the rotor speed to be estimated, $J$ is the rotor inertia, $T_L$ is the load torque, and $y$ is the output vector composed of the stator currents. The rest symbols are parameters, where $\sigma=L_s(1-L_m^3 / L_s L_r)$, $\alpha=R_r/L_r$, $\beta = L_m/\sigma L_r$, $\gamma=R_s/\sigma+\alpha\beta L_m$, $\mu=3L_m/2L_r$; $( R_s, L_s)$ and $(R_r, L_r)$ are the resistance-inductance pairs of the stator and rotor, respectively; $L_m$ is the mutual inductance. As shown above, the state vector $x$ comprises $i_{ds}$, $i_{qs}$, $\psi_{dr}$, $\psi_{qr}$, and $\omega$. The parameter setting follows~\cite{Zhou:ACC:2016}. Note that, because of the focus on state estimation, an open-loop control scheme is considered with $u_{ds}(t)=380 \sin(100\pi t)$ and $u_{qs}(t)=-380 \sin(200 \pi t)$. The state estimation problem is then to estimate the entire state vector through time using the measurement data of $i_{ds}$, $i_{qs}$, $u_{ds}$ and $u_{qs}$. 

In the simulation, the model is initialized with $x_0= [0 \ 0 \ 0 \ 0 \ 0]^\T$. The initial state guess for all the filters is set to be $\hat x_{0|0}= [1 \ 1 \ 1 \ 1 \ 1]^\T$ and  $P_0^x = 10^{-2} I$. For EnKF, its estimation accuracy depends on the ensemble size. Thus, different sizes are implemented to examine this effect, with $N_s= 40$, 100, 200 and 400. To make a fair evaluation, EnKF with each $N_s$ is run for 100 times as a means to reduce the influence of randomly generated noise. The estimation error for each run is defined as $\sum_k \|x_k -\hat x_{k|k} \|_2$; the errors from the 100 runs are averaged to give the final estimation error for comparison. 

Fig.~\ref{Estimation-error} shows the estimation errors for $\omega$ along with $\pm 3 \sigma$ bounds in a simulation run of EKF, UKF and EnKF with ensemble size of 100 (here, $\sigma$ stands for the standard deviation associated with the estimate of $\omega$, and $\pm 3 \sigma$ bounds correspond to the 99\% confidence region). It is seen that, in all three cases, the error is large at the initial stage but gradually decreases to a much lower level, indicating that the filters successfully adapt their running according to their own equations. In addition, UKF demonstrates the best estimation of $\omega$ overall. The average estimation errors over 100 runs are summarized in Table~\ref{EKF-UKF-EnKF-Error}. It also shows that UKF offers the most accurate estimation when all state variables are considered. In addition, the estimation accuracy of EnKF improves when the ensemble size increases.

\begin{table*}[tp]\color{blue}
\centering
\caption{Comparison of EKF, UKF and EnKF.}
\label{EKF-UKF-EnKF-Comparison}
\begin{tabular}{c|c|c|c|c}\hline
     & \parbox{0.12\linewidth}{\centering Computational complexity} & Jacobian matrix & System dimensions & Applications                                                                                                                                                                                                                                                                                                                                             \\\hline
\multirow{4}{*}{EKF}&\multirow{4}{*}{High}&\multirow{4}{*}{Needed}&\multirow{4}{*}{Low}& \multirow{8}{*}{\parbox{0.4\linewidth}{Guidance and navigation, flight control,  attitude control, target tracking, robotics (e.g., simultaneous localization and mapping), electromechanical systems (e.g., induction motors and electric drives), vibration control, biomedical signal processing, sensor fusion, structural system monitoring, sensor networks, process control, computer vision, battery management, HVAC systems, %(heating, ventilation and air conditioning)  
econometrics }}
\\ &&&& 
\\ &&&& \\ &&&& \\\cline{1-4} \multirow{4}{*}{UKF}&\multirow{4}{*}{High}&\multirow{4}{*}{Not needed}&\multirow{4}{*}{Low to medium}
\\ &&&& 
\\ &&&& 
\\ &&&&\\\hline
EnKF & Low                      & Not needed      & High              &  \parbox{0.4\linewidth}{Meteorology, hydrology, weather forecasting, oceanography, reservoir engineering, transportation systems, power systems}\\\hline
\end{tabular}
\end{table*}

We draw the following remarks about nonlinear state estimation from our extensive simulations with this specific problem and experience with other problems in our past research.
\begin{itemize}

\item The initial estimate can significantly impact the estimation accuracy. For the problem considered here, it is found that EKF and EnKF are more sensitive to an initial guess. It is noteworthy that an initial guess, if differing much from the truth, can lead to divergence of filters. Hence, one is encouraged to obtain a guess as close as possible to the truth through using prior knowledge or trying different guesses.

\item A filter's performance can be problem-dependent. A filter can provide estimation at a decent accuracy when applied to a problem but may fail when used to handle another. Thus, the practitioners are suggested to try different filters whenever allowed to find out the one that performs the best for his/her specific problem. 

\item Successful application of a filter usually requires to tune the covariance matrices and in some cases, parameters involved in a filter (e.g., $\lambda$, $\alpha$ and $\beta$ in UKF), because of their important influence on estimation~\cite{Ge:TAC:2016}. The trial-and-error method is common in practice. Meanwhile, there also exist some studies of systematic tuning methods, e.g.,~\cite{Scardua:AUTO:2017,Scala:TSP:1996}. Readers may refer to them for further information.

\item In choosing the best filter, engineers need to take into account all the factors relevant to the problem they are addressing, including but not limited to estimation accuracy, computational efficiency, system's structural complexity, and problem size. To facilitate such a search, Table~\ref{EKF-UKF-EnKF-Comparison} summarizes the main differences and application areas of EKF, UKF and EnKF.

\end{itemize}

}
\section{Other Filtering Approaches and Estimation Problems}\label{Other-Problems}

Nonlinear stochastic estimation remains a major research challenge for the control research community.
Continual research effort has been in progress toward the development of advanced methods and theories in addition to the  KFs reviewed above. This section gives an overview of other major filtering approaches.

{\bf Gaussian filters (GFs).} GFs are a class of Bayesian filters enabled by a series of Gaussian distribution approximations. They bear much resemblance with KFs in view of their prediction-update structure and thus, in a broad sense, belong to the KF family. As seen earlier, the KF-based estimation relies on the evaluation of a set of integrals indeed---for example, the prediction of $x_k$ is attained in~\eqref{GF_P_1} by computing the conditional mean of $f(x_{k-1})$ on $\bbY_{k-1}$. The equation is repeated here for convenience of reading:
\begin{align*}
\hat x_{k|k-1} &= \E\left[ f(x_{k-1} ) | \bbY_{k-1} \right]\\&= \int f(x_{k-1}) p(x_{k-1} | \bbY_{k-1}) \rmd x_{k-1}.
\end{align*}
GFs approximate $p(x_{k-1} | \bbY_{k-1})$ with a Gaussian distribution having mean $\hat x_{k-1|k-1}$ and covariance $P_{k-1|k-1}^x$. Namely, $p(x_{k-1} | \bbY_{k-1})$ is replaced by $\calN(\hat x_{k-1|k-1}, P_{k-1|k-1}^x )$~\cite{Ito:TAC:2000}. Continuing with this assumption, one can use the {\em Gaussian quadrature integration rules} to evaluate the integral. A quadrature is a means of approximating a definite integral of a function by a weighted sum of values obtained by evaluating the function at a set of deterministic points in the domain of integration. An example of a one-dimensional Gaussian quadrature is the Gauss-Hermite quadrature, which plainly states that,
for a given function $g(x)$,
\begin{align*}
\int_{-\infty}^\infty g(x)\cdot\calN(x;0,1) \rmd x \approx \sum_{i=1}^m w_i g(x_i),
\end{align*}
where $m$ is the number of points used, $x_i$ for $i=1,2,\cdots,m$ the roots of the Hermite polynomial $H_m(x)$, and $w_i$ the associated weights
\begin{align*}
w_i = \frac{2^{m-1}m!\sqrt{\pi}}{m^2\left[ H_{m-1}(x_i)\right]^2}.
\end{align*}
Exact equality holds for polynomials of order up to $2m-1$. Applying the multivariate version of this quadrature, one can obtain a filter in a KF form, which is named {\em Gauss-Hermite KF} (GHKF)~\cite{Ito:TAC:2000,Jia:JGCD:2011}. GHKF reduces to  UKF in certain cases~\cite{Ito:TAC:2000}. Besides, the cubature rules for numerical integration can also be used in favor of a KF realization, which  yields a {\em cubature Kalman filter} (CKF)~\cite{Arasaratnam:TAC:2009, Jia:AUTO:2013}. It is noteworthy that CKF is a special case of UKF given $\alpha = 1$, $\beta = 0$ and $\kappa = 0$~\cite{Sarkka:axXiv:2015}.

{\bf Gaussian-sum filters (GSFs).} Though used widely in the development of GFs and KFs, Gaussianity approximations are often inadequate and performance-limiting for many systems. To deal with a non-Gaussian pdf, GSFs represent it by a weighted sum of Gaussian basis functions~\cite{Anderson:1979}. For instance, the {\em a posteriori} pdf of $x_k$ is approximated by
\begin{align*}
p(x_k | \bbY_K) = \sum_{i=1}^m W_k^i \calN \left(x_k; \hat x_{k|k}^i, P_{k|k}^i\right),
\end{align*}
where $W_k^i$, $\hat x_k^i$ and $P_{k|k}^i$ are the weight, mean and covariance of the $i$-th Gaussian basis function (kernel), respectively. This can be justified by the Universal Approximation Theorem, which states that a continuous function can be approximated by a group of Gaussian functions with arbitrary accuracy under some conditions~\cite{Park::NC:1991}. A GSF then recursively updates $\hat x_{k|k}^i$, $P_{k|k}^i$ and $W_k^i$. In the basic form, $\hat x_{k|k}^i$ and $P_{k|k}^i$ for $i=1,2,\cdots,m$ are updated individually through EKF, and $W_k^i$ updated according to the output-prediction accuracy of $\hat x_k^i$.
The assumption for the EKF-based update is that the system's nonlinear dynamics can be well represented by aggregating linearizations around a sufficient number of different points (means).
In recent years, more sophisticated GSFs have been developed by merging the Gaussian-sum pdf approximation with other filtering techniques such as UKF, EnKF, GFs and particle filtering~\cite{Straka:Fusion:2011,Dovera:CG:2011,Ito:TAC:2000,Horwood:TAC:2011,Kotecha:TSP:2003} or optimization techniques~\cite{Terejanu:TAC:2011}.

\begin{figure}[t]
  \center
  \includegraphics[width=\linewidth]{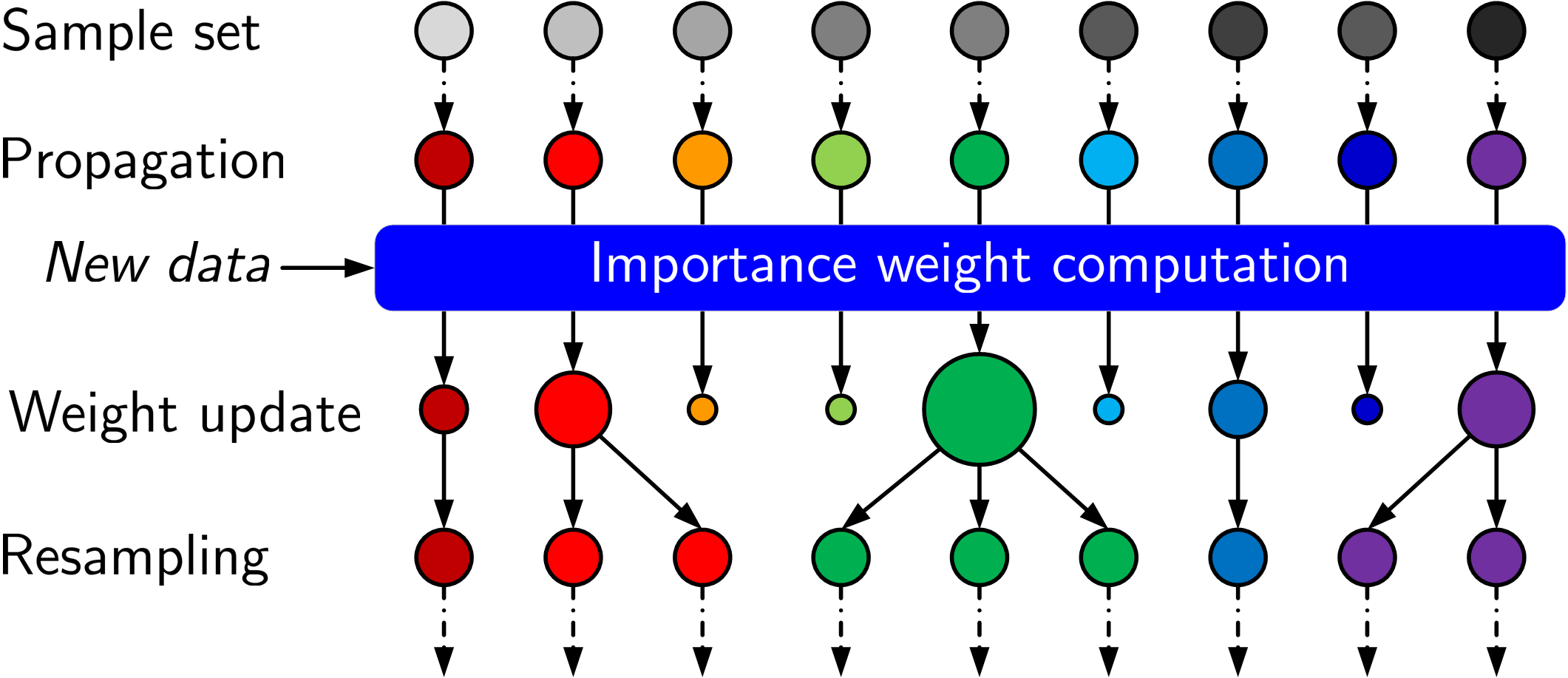}\\
  \caption{A graphic diagram of the PF technique modified from~\cite{Parker:STR:2013}. Suppose that a set of samples (particles, as shown in gray color in the figure) are used to approximate the conditional pdf of the state on the available measurements as a particle discrete distribution. A one-step-froward propagation is implemented to generate the samples for the state at the next time instant. On its arrival, the new measurement will be used to update the weight of each sample to reflect its importance relative to others. Some samples may be given almost zero weight, referred to as degeneracy, and thus have meaningless contribution to the state estimation. Resampling will then be performed to generate a new set of samples.}\label{Particle-Filter}
\end{figure}

{\bf Particle filters (PFs).} The PF approach was first proposed in the 1990s~\cite{Gordon:RSP:1993} and came to prominence soon after that owing to its capacity for high-accuracy nonlinear non-Gaussian estimation. Today they have grown into a broad class of filters. As random-sampling-enabled numerical realizations of the Bayesian filtering principle, they are also known as the sequential Monte Carlo methods in the literature. Here, we introduce the essential idea with minimum statistical theory to offer the reader a flavor of this approach. Suppose that $N_s$ samples, $\hat x_{k-1|k-1}^{(i)}$ for $i=1,2,\cdots,N_s$ are drawn from $p(x_{k-1}| \bbY_{k-1})$ at time $k-1$. The $i$-th sample is associated with a weight $W_{k-1}^{(i)}$, and $\sum_{i=1}^{N_s} W_{k-1}^{(i)}=1$. Then, $p(x_{k-1}|\bbY_{k-1})$ can be empirically described as
\begin{align}\label{Empirical}
p(x_{k-1} | \bbY_{k-1}) \approx  \sum_{i=1}^{N_s} W_{k-1}^{(i)} \delta \left(x_{k-1} - \hat x_{k-1|k-1}^{(i)}\right).
\end{align}
This indicates that the estimate of $x_{k-1}$ is
\begin{align*}
\hat x_{k-1|k-1} &= \int x_{k-1} p(x_{k-1}|\bbY_{k-1}) \rmd x_{k-1}\\& = \sum_{i=1}^N W_{k-1}^{(i)} \hat x_{k-1|k-1}^{(i)}.
\end{align*}
The samples can be propagated one-step forward to generate a sampling-based description of $x_k$, i.e.,
\begin{align*}
\hat x_{k}^{(i)} = f\left( \hat x_{k-1|k-1}^{(i)} \right) + w_{k-1}^{(i)}, \ i =1,2,\cdots,N_s,
\end{align*}
where $w_{k-1}^{(i)}$ for $i=1,2,\cdots,N_s$ are samples drawn from the distribution of $w_{k-1}$.
After the propagation, each new sample should take a different weight in order to be commensurate with its probabilistic importance with respect to the others. To account for this, one can evaluate $p\left(y_k | \hat x_{k}^{(i)} \right)$, which quantifies the likelihood of $y_k$ given the $i$-th sample $\hat x_k^{(i)}$. Then, the weight can be updated and normalized on $[0,1]$ by
\begin{align*}
W_k^{(i)} = W_{k-1}^{(i)} p\left(y_k | \hat x_{k}^{(i)} \right), \ W_k^{(i)} = \frac{W_k^{(i)}}{\sum_{i=1}^{N_s} W_k^{(i)}}.
\end{align*}
Then, an empirical sample-based distribution is built for $p(x_k|\bbY_k)$ as in~\eqref{Empirical}, and the estimate of $x_k$ can be computed as
\begin{align*}
\hat x_k &= \sum_{i=1}^{N_s} W_k^{(i)} \hat x_k^{(i)}.
\end{align*}
In practical implementation of the above procedure, the issue of {\em degeneracy} may arise, which refers to the scenario that many or even most samples take almost zero weights. Any occurrence of degeneracy renders the affected samples useless. Remedying this situation requires the deployment of {\em resampling}, which replaces the samples by new ones drawn from the discrete empirical distribution defined by the weights. Summarizing the steps of sample propagation, weight update and resampling gives rise to a basic PF, which is schematically shown in Fig.~\ref{Particle-Filter}. While the above outlines a reasonably intuitive explanation of the PF approach, a rigorous development can be made on a solid statistical foundation, as detailed in~\cite{Sarkka:CamPress:2013,Doucet:Ocford:2011,Arulampalam:TSP:2002}.

\begin{figure*}
\begin{align}\label{Bayesian_MHE}\nonumber
\{ \hat x_{k-N}, \cdots, \hat x_k\} &= \arg\max_{x_{k-N},\cdots,x_k} p(x_{k-N}, \cdots, x_k | y_{k-N}, \cdots, y_k)\\\nonumber
&= \arg\max_{x_{k-N},\cdots,x_k} p(x_{k-N}) \prod_{l = k-N}^{k} p(y_l | x_l) \prod_{l = k-N}^{k-1} p(x_{l+1} | x_l)\\
&= \arg\max_{x_{k-N},\cdots,x_k} p(x_{k-N}) \prod_{l = k-N}^{k} p(v_l) \prod_{l = k-N}^{k-1}p(w_l).
\end{align}
\hrulefill
% The spacer can be tweaked to stop underfull vboxes.
\vspace*{4pt}
\end{figure*}

%Convergence~\cite{Crisan:TSP:2002}
With the sample-based pdf approximation, PFs can demonstrate estimation accuracy superior to other filters given a sufficiently large $N_s$. It can be proven that their estimation error bound does not depend on the dimension of the system~\cite{Crisan:TechReport:2000}, implying applicability for high-dimensional systems. A possible limitation is their computational complexity, which comes at $O(N_s n_x^2)$ with $N_s\gg n_x$. Yet, a strong anticipation is that the rapid growth of computing power tends to overcome this limitation, enabling wider application of PFs. A plethora of research has also been undertaken toward computationally efficient PFs~\cite{Yi:TSP:2013}. A representative means is the Rao-Blackwellization that applies the standard KF to the linear part of a system and a PF to the nonlinear part and reduces the number of samples to operate on~\cite{Sarkka:CamPress:2013}. The performance of PFs often relies on the quality of samples used. To this end, KFs can be used in combination to provide high-probability particles for PFs, leading to a series of combined KF-PF techniques~\cite{Freitas:NC:2000,Merwe:NIPS:2001,Li:ISCIT:2005}.
A recent advance is the implicit PF, which uses the implicit sampling method to generate samples capable of an improved approximation of the pdf~\cite{Chorin:PNSA:2009,Chorin:CAMCS:2010}.

%{\bf Particle filters.}

\begin{figure}[t]
  \centering
  \includegraphics[width=0.7\linewidth]{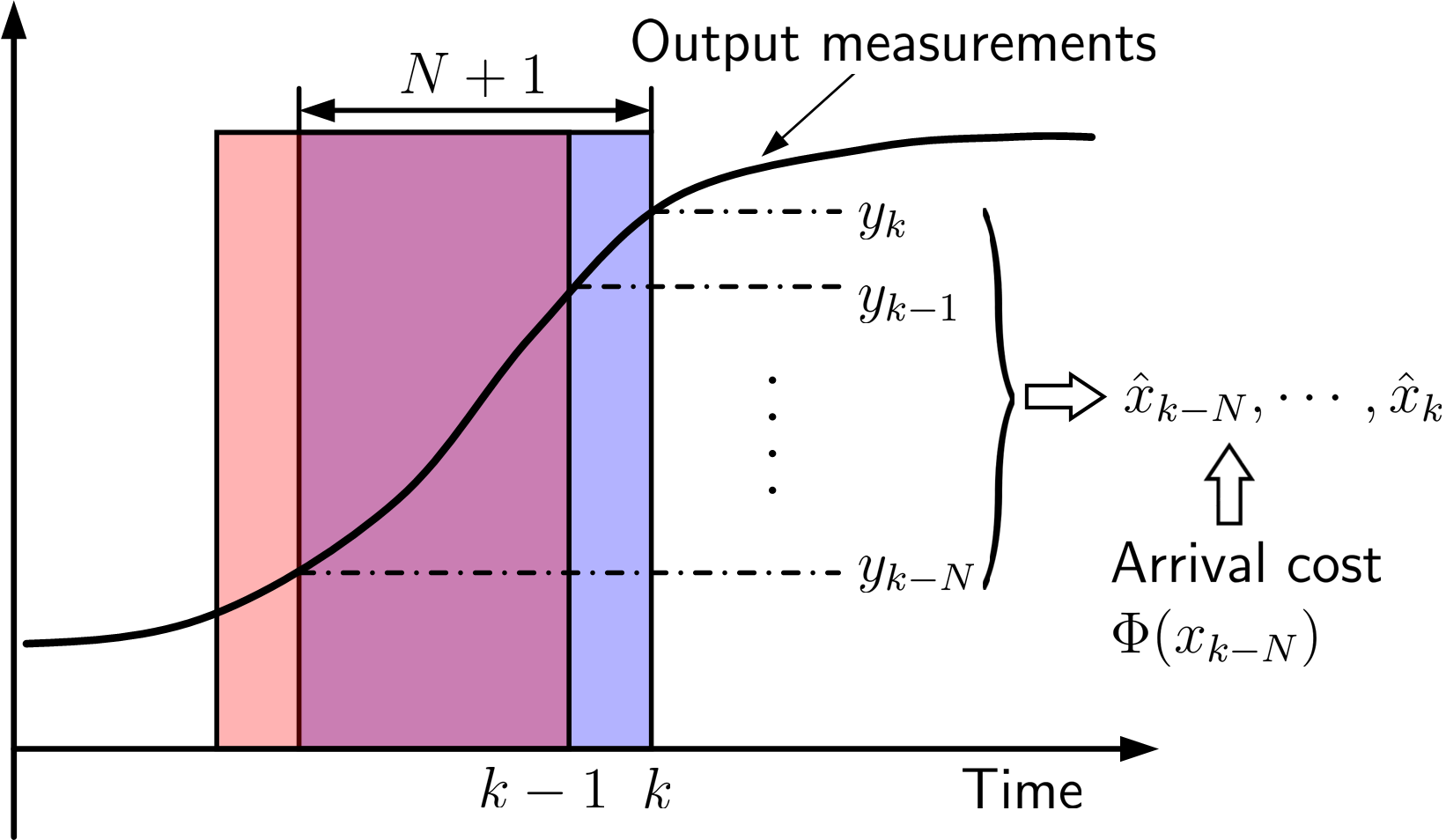}
  \caption{MHE performed over a moving observation horizon that spans $N+1$ time instants. For estimation at time $k$, the arrival cost $\Phi(x_{k-N})$ is determined first, which summarizes the information of the system behavior up to the beginning of the horizon. Then, the output measurements within the horizon, $y_{k-N},\cdots,y_k$,  are used, along with the arrival cost, to conduct estimation of $\hat x_{k-N}, \cdots, \hat x_k$ through constrained optimization.}\label{MHE-Figure}
\end{figure}

{\bf Moving-horizon estimators (MHEs).} MHEs are an emerging estimation tool based on constrained optimization. In general, they aim to find the state estimate through minimizing a cost function subject to certain constraints. The cost function is formulated on the basis of the system's behavior in a moving horizon. To demonstrate the idea, we consider the Maximum a Posteriori estimation (MAP) for the system in~\eqref{sys} during the horizon $[k-N,k]$ as shown in~\eqref{Bayesian_MHE}. Assuming $w_k \sim \calN(0,Q)$ and $v_k \sim \calN(0,R)$ and using the logarithmic transformation, the above cost function becomes
\begin{align*}
\min_{x_{k-N},\cdots,x_k} &\Phi(x_{k-N}) + \sum_{l=k-N}^{k-1}w_l^\T Q^{-1} w_l +\sum_{l=k-N}^k v_l^\T R^{-1} v_l,
\end{align*}
where $\Phi(x_{k-N})$ is the arrival cost summarizing the past information up to the beginning of the horizon. The minimization here should be subject to the system model in~\eqref{sys}. Meanwhile, some physically motivated constraints for the system behavior should be incorporated. This produces the formulation of MHE given as
\begin{align*}
\{ \hat x_{k-N}, \cdots, \hat x_k\} &= \argmin_{x_{k-N},\cdots,x_k} \Phi(x_{k-N}) \\&  + \sum_{l=k-N}^{k-1}w_l^\T Q^{-1} w_l +\sum_{l=k-N}^k v_l^\T R^{-1} v_l,\\
{\rm subject \ to:} & \quad x_{l+1} = f(x_l)+w_l,\\
& \quad y_l = h(x_l) + v_l,\\
&  \quad x \in {\mathbb X}, \ w \in {\mathbb W}, \ v \in {\mathbb V},
\end{align*}
where ${\mathbb X}$, ${\mathbb W}$ and ${\mathbb V}$ are, respectively, the sets of all feasible values for $x$, $w$ and $v$ and imposed as the constraints.
It is seen that MHE tackles the state estimation through constrained optimization executed over time in a receding-horizon manner, as shown in Fig.~\ref{MHE-Figure}. For an unconstrained linear system, MHE reduces to the standard KF. It is worth noting that the arrival cost $\Phi(x_{k-N})$ is crucial for the performance or even success of MHE approach. In practice, an exact expression is often unavailable, thus requiring an approximation~\cite{Rao:TAC:2003,Rao:Thesis:2000}. With the deployment of constrained optimization, MHE is computationally expensive and usually more suited for slow dynamic processes; however, the advancement of real-time optimization has brought some promises to its faster implementation~\cite{Tenny:ACC:2002, Diehl:Springer:2009}.

{\bf Simultaneous state and parameter estimation (SSPE).} In state estimation problems, a system model is considered fully known {\em a priori}. This may not be true in various real-world situations, where part or even all of the model parameters are unknown or subject to time-varying changes. Lack of knowledge of the parameters can disable an effort for state estimation in such a scenario. Hence, SSPE is motivated to enable state estimation self-adapting to the unknown parameters. Despite the complications, a straightforward and popular way for SSPE is through state augmentation. To deal with the parameters, the state vector is augmented to include them, and on account of this, the state-space model will be transformed accordingly to one with increased dimensions. Then, a state estimation technique can be applied directly to the new model to estimate the augmented state vector, which is a joint estimation of the state variables and parameters. In the literature, EKF, UKF, EnKF and PFs have been modified using this idea for a broad range of applications~\cite{Ching:PEM:2006,Fang:ACC:2013,Fang:CEP:2014,Chowdhary:AST:2010,Chen:JPC:2005}. Another primary solution is the so-called {\em dual Kalman filtering}. By ``dual'', it means that the state estimation and parameter estimation are performed in parallel and alternately. As such, the state estimate is used to estimate the parameters, and the parameter estimate is used to update the state estimation. Proceeding with this idea, EKF, UKF and EnKF can be dualized~\cite{Wan:Wiley:2001b,Wan:NIPS:2000,Vandyke:AAS:2004,Moradkhani:AWR:2005}. It should be pointed out that caution should be taken when an SSPE approach is developed. Almost any SSPE problem is nonlinear by nature with coupling between state variables and parameters. The joint state observability and parameter identifiability may be unavailable, or the estimation may get stuck in local minima. Consequently, the convergence can be vulnerable or unguaranteed, diminishing the chance of successful estimation. Thus application-specific SSPE analysis and development are recommended.

{\bf Simultaneous state and input estimation (SSIE).} Some practical applications entail not only unknown states but also unknown inputs. An example is the operation monitoring for an industrial system subject to unknown disturbance, where the operational status is the state and the disturbance the input. In maneuvering target tracking, the tracker often wants to estimate the state of the target, e.g., position and velocity, and the input, e.g., the acceleration. Another example is the wildfire data assimilation extensively investigated in the literature. The spread of wildfire is often driven by local meteorological conditions such as the wind. This gives rise to the need for a joint estimation of both the fire perimeters (state) and the wind speed (input) toward accurate monitoring of the fire growth.

The significance of SSIE has motivated a large body of work. A lead was taken in~\cite{Mendel:TAC:77} with
the development of a
KF-based approach to estimate the state and external white process noise
of a linear discrete-time system~\cite{Mendel:TAC:77}. Most recent research builds on the existing state estimation
techniques. Among them, we highlight
KF~\cite{Hsieh:TAC:2000,Hsieh:AJC:2010}, MHE~\cite{Pina:AUTO:2006},
$H_\infty$-filtering~\cite{You:AUTOSinica:2008}, sliding mode
observers~\cite{Floquet:IJACSP:2007,Bejarano:IJRNC:2007}, and
minimum-variance unbiased estimation~\cite{Gillijns:AUTO:2007:b,Gillijns:AUTO:2007:a, Fang:ACC:2008, Fang:ACSP:2011,Fang:AUTO:2012}. SSIE for nonlinear systems involves more complexity, with fewer results reported. In~\cite{Corless:AUTO:1998,Ha:AUTO:2004}, SSIE is investigated for a special class of
nonlinear systems that consist of a nominally linear part and a nonlinear
part. However, the Bayesian statistical thinking has been generalized to address this topic, exemplifying its power in the development of nonlinear SSIE approaches.
In~\cite{Fang:AUTO:2013,Fang:ACSP:2015}, a Bayesian approach along with numerical optimization is taken to achieve SSIE for nonlinear systems of a general form. This Bayesian approach is further extended in~\cite{Fang:CDC:2013,Fang:CEP:2017} to build an ensemble-based SSIE method, as a counterpart of EnKF, for high-dimensional nonlinear systems. It is noteworthy that SSIE and SSPE would overlap if we consider the parameters as a special kind of inputs to the system. In this case, the SSIE approaches may find their use in solving SSPE problems.

\section{Conclusion}
%Theoretical: Robustness and Applicability.
This article offered a state-of-the-art review of nonlinear state estimation approaches.
As a fundamental problem encountered across a few research areas, nonlinear stochastic estimation has stimulated a sustaining interest during the past decades. In the pursuit of solutions, the Bayesian analysis has proven to be a time-tested and powerful methodology for addressing various types of problems. %, thanks to its capability of updating the probabilistic belief in unknown quantities by using new evidence. 
In this article, we first introduced the Bayesian thinking for nonlinear state estimation, showing the nature of state estimation from the perceptive of Bayesian update. Based on the notion of Bayesian state estimation, a general form of the celebrated KF is derived. Then, we illustrated the development of the standard KF for linear systems and EKF, UKF and EnKF for nonlinear systems. A case study of state estimation for speed sensorless induction motors was provided to present a comparison of the EKF, UKF and EnKF approaches. We further extended our view to a broader horizon including GF, GSF, PF and MHE approaches, which are also deeply rooted in the Bayesian state estimation and thus can be studied from a unified Bayesian perspective to a large extent. 

{\color{blue}
Despite remarkable progress made thus far, it is anticipated that nonlinear Bayesian estimation continues to see intensive research in the coming decades. This trend will be partially driven by the need to use state estimation as a mathematical tool to enable various emerging systems in contemporary industry and society, stretching from autonomous transportation to sustainable energy and smart X (grid, city, planet, geosciences, etc.). Here, we envision several directions that may shape the future research in this area. The first one lies in accurately characterizing the result of a nonlinear transformation applied to a probability distribution. Many of the present methods such as EKF, UKF and EnKF were more or less motivated to address this fundamental challenge. However, there still exists no solution generally acknowledged as being satisfactory, leaving room for further exploration. Second, much research is needed to deal with uncertainty. Uncertainty is intrinsic to many practical systems because of unmodeled dynamics, external disturbances, inherent variability of a dynamic process, and sensor noise, representing a major threat to successful estimation. Although the literature contains many results on state estimation with robustness to uncertainty, the research has not reached a level of maturity because of the difficulty involved. A third research direction is optimal sensing structure design. Sensing structure or sensor deployment is critical for data informativeness and thus can significantly affect the effectiveness of estimation. An important question thus is how to achieve co-design of a sensing structure and Bayesian estimation approach to maximize estimation accuracy. Fourth, Bayesian estimation in a cyber-physical setting is an imperative. Standing at the convergence of computing, communication and control, cyber-physical systems (CPSs) are foundationally important and underpinning today's smart X initiatives. They also present new challenges for estimation, which include communication constraints or failures, computing limitations, and cyber data attacks. The current research is particularly rare on nonlinear Bayesian estimation for CPSs. Finally, many emerging industrial and social applications are data-intensive, thus asking for a seamless integration of Bayesian estimation with big data processing algorithms. New principles, approaches and computing tools must be developed to meet this pressing need, which will make an unprecedented opportunity to advance the Bayesian estimation theory.
}

\renewcommand{\theequation}{A.\arabic{equation}}
\setcounter{equation}{0}
\appendix
This appendix offers a summary of the properties of the Gaussian distribution.
Suppose $z\in \bbR^n$ is a Gaussian random vector with $z\sim \calN (\bar z, P_z)$. The pdf of $z$ is  expressed as
\begin{align*}
p(z) = \frac{1}{(\sqrt{(2\pi)^n |P_z|}} \exp\left(- (z-\bar z)P_z^{-1}(z-\bar z)^\T \right).
\end{align*}
Some useful properties of the Gaussian vectors are as follows~\cite{Candy:2009}.
\begin{enumerate}%[label={Property \arabic*)},leftmargin=5.6em]

\item 
\begin{align}\label{GRV_1} \nonumber
&\int z p(z) \rmd z = \bar z, \ \int (z-\bar z)(z-\bar z)^\T p(z) \rmd z = P_z,  \\ & \int zz^\T p(z) \rmd z =P_z+\bar z \bar z^\T.
\end{align}

\item The affine transformation of $z$, $Az+b$, is Gaussian, i.e.,
\begin{align}\label{GRV_2}
Az+b \sim \calN \left(A\bar z+b, AP_zA^\T \right).
\end{align}

\item The sum of two independent Gaussian random vectors is Gaussian; i.e., if $z_1\sim \calN (\bar z_1, P_{z_1})$ and $z_2\sim \calN (\bar z_2, P_{z_2})$ and if $z_1$ and $z_2$ are independent, then
\begin{align}\label{GRV_3}
Az_1+Bz_2 \sim \calN \left(A \bar z_1+ B \bar z_2, A P_{z_1}A^\T+ BP_{z_2}B^\T \right).
\end{align}

\item For two random vectors jointly Gaussian, the conditional distribution of one given the other is Gaussian. Specifically, if $z_1$ and $z_2$ are jointly Gaussian with
\begin{align*}
\left[
\begin{matrix}
z_1 \cr z_2
\end{matrix}
\right] \sim \calN \left( \left[
\begin{matrix}
\bar z_1 \cr \bar z_2
\end{matrix}
\right], \left[
\begin{matrix}
P_{z_1} & P_{z_1z_2} \cr P_{z_1z_2}^\T & P_{z_2}
\end{matrix}
\right] \right),
\end{align*}
then
\begin{align}\label{GRV_4} \nonumber
z_1|z_2 &\sim \calN \Big(\bar z_1 + P_{z_1z_2}P_{z_2}^{-1}(z_2-\bar z_2), \\  & \quad\quad\quad  P_{z_1}-P_{z_1z_2} P_{z_2}^{-1}P_{z_1z_2}^\T \Big).
\end{align}
\end{enumerate}

\balance
{\small
\bibliographystyle{IEEEtran}
%\bibliography{Stochastic_Estimation,Battery_SOC}
\bibliography{Stochastic_Estimation}
}

\end{document}